\documentclass[lettersize,journal]{IEEEtran}
\usepackage{amsmath,amsfonts}
\usepackage{algorithmic}
\usepackage{algorithm}
\usepackage{array}
\usepackage[caption=false,font=normalsize,labelfont=sf,textfont=sf]{subfig}
\usepackage{textcomp}
\usepackage{color}
\usepackage{stfloats}
\usepackage{url}
\usepackage{cite}
\usepackage{verbatim}
\usepackage{graphicx}
\usepackage{amsthm}
\begin{document}

\title{Consensus Formation Tracking for Multiple AUV Systems Using Distributed Bioinspired\\ Sliding Mode Control}

\author{Tao Yan,~\IEEEmembership{Student Member,~IEEE}, Zhe Xu,~\IEEEmembership{Student Member,~IEEE}, Simon X. Yang,~\IEEEmembership{Senior Member,~IEEE,}
        % <-this % stops a space
% \thanks{This paper was produced by the IEEE Publication Technology Group. They are in Piscataway, NJ.}% <-this % stops a space
\thanks{This work was supported by the Natural Sciences and Engineering Research Council (NSERC) of Canada.}
\thanks{The authors are with Advanced Robotics and Intelligent Systems (ARIS)
Laboratory, School of Engineering, University of Guelph, Guelph, ON.
N1G2W1, Canada (e-mail: tyan03; zxu02; syang@uoguelph.ca).}
\thanks{Corresponding author: Simon X. Yang }}

% The paper headers
% \markboth{Journal of \LaTeX\ Class Files,~Vol.~14, No.~8, August~2021}%
% {Shell \MakeLowercase{\textit{et al.}}: A Sample Article Using IEEEtran.cls for IEEE Journals}

% \IEEEpubid{0000--0000/00\$00.00~\copyright~2021 IEEE}
% Remember, if you use this you must call \IEEEpubidadjcol in the second
% column for its text to clear the IEEEpubid mark.

\maketitle

\begin{abstract}
Consensus formation tracking of multiple autonomous underwater vehicles (AUVs) subject to nonlinear and uncertain dynamics is a challenging problem in robotics. To tackle this challenge, a distributed bioinspired sliding mode controller is proposed in this paper. First, the conventional sliding mode controller (SMC) is presented, and the consensus problem is addressed on the basis of graph theory. Next, to tackle the high frequency chattering issue in SMC scheme and meanwhile improve the robustness to the noises, a bioinspired approach is introduced, in which a neural dynamic model is employed to replace the nonlinear sign or saturation function in the synthesis of conventional sliding mode controllers. Furthermore, the input-to-state stability of the resulting closed-loop system is proved in the presence of bounded lumped disturbance by the Lyapunov stability theory. Finally, simulation experiments are conducted to demonstrate the effectiveness of the proposed distributed formation control protocol.
\end{abstract}

\begin{IEEEkeywords}
Bioinspired approach, consensus formation tracking, distributed sliding mode control, graph theory, multiple AUVs.
\end{IEEEkeywords}

\section{Introduction}
\IEEEPARstart{A}{utonomous} underwater vehicles (AUVs) are a class of marine mechatronic systems capable of operating underwater automatically for a long period of time. This type of machine enables human beings to have the abilities to explore and exploit the marine resources. A great number of applications both of commercial and military purposes are surged by employing such vehicles, including oceanographic seabed mapping, gas and oil exploration, payload delivery, etc. Thus, developing high performance AUV systems has been a pressing topic in both control and ocean engineering communities \cite{Li,1,3,4,25,zhu2020novel}.

In recent decades, the number of AUVs is becoming larger in order to adapt to more complex marine missions, such as large-scale mine sweeping, undersea sampling and mapping, oil and gas exploration, etc. {In addition to that, multiple autonomous vehicle systems are proven to be able to collect more robust data and exhibit more fault-tolerant capabilities than the single vehicles \cite{5}. }Therefore, the coordination problem among multiple AUVs has attracted much attention. Particularly, the formation control as the fundamental problem in multiple vehicle coordination has been widely studied \cite{6}. However, due to the nonlinear and uncertain nature of AUV's characteristics as well as the complexity of underwater situations, {steering a fleet of unmanned marine vehicles to keep a desired formation configuration precisely may be  in front of many challenges \cite{lakhekar2019disturbance,wang2020active}. }

A large amount of work has been done to develop an effective formation control protocol for a team of AUVs. {The coordination strategies can typically be classified as behavior-based approach, virtual structure, leader-follower structure, and artificial potential approaches.} In behavioral approaches \cite{6,8}, decentralized control implementations are used to design local controllers for each individual, where the control law is of the weighted form to meet various control objectives, such as formation keeping, goal seeking, path following, etc. {Because of the separative feature of such approaches, it provides a more flexible and natural way to synthesize controllers to satisfy the collective behaviors.} Nonetheless, the main drawback relies on the fact that quite often it is hard to formulate the problem and provide the theoretical guarantees. In virtual structure \cite{9,10}, the desired formation shape is viewed as a single rigid body, and the reference points are derived for individuals according to the overall formation, by which the coordination problem could be cast into a series of trajectory tracking problems for each vehicle. Although this facilitates the design of formation controllers, the cooperative performance is greatly sacrificed as a whole. To improve cooperation among vehicles, the leader-follower structure is proposed, where one or several vehicle is selected as the leader which is able to access the desired formation and reference trajectory, and the rest of the vehicles are treated as the followers with the sole objective to maintain the desired posture (position and orientation) to the leader \cite{11,12}. Due to its simplicity as well as the scalability to a larger group,  {such a structure has been extensively employed in multi-robot system's formation control. Another class of widely used approach in multiple vehicle cooperation is referred to as the artificial potential field, first developed to address the problems of goal seeking as well as collision avoiding, and then adapted to the formation control by designing proper potential functions for individuals which may produce the corresponding potential fields enforcing the vehicles to maintain a relative pose with their neighbors \cite{13,14}.} Nevertheless, the biggest issue of such approach is that it has chance to enter into the local minimum, which may render the entire system malfunction.

In addition to the coordination strategies, there also are many of efforts being put to improve the performance of motion control of AUVs in view of strong nonlinear characteristics as well as complicated marine environments. The robust optimal formation control for a fleet of marine vehicles subject to communication delays was addressed based on the leader-follower structure \cite{15}. Adaptive robust formation control of AUVs was studied to handle the time-varying environmental disturbances \cite{16}. An adaptive sliding mode formation control scheme was proposed based on the communication consensus to allow for the variable added mass as well as the communication constraints \cite{22}. {To tackle the chattering in sliding mode control, a novel reaching law was designed, in which an adaptive gain was incorporated based on the variation of the sliding surface \cite{ramezani2019novel}. Furthermore, taking into account the unavailable velocity measure as well as reduction of the conservativeness, a equivalent output injection adaptive sliding mode observer based terminal sliding mode control approach was proposed for trajectory tracking of underwater vehicles \cite{wang2016multivariable}.} Formation control for underactuated AUVs was addressed based on the leader-follower backstepping scheme; moreover, both parametric uncertainty and unknown disturbance arising in dynamic model were handled \cite{23}. {To further improve the formation accuracy, neural adaptive formation control for multi-underactuated marine vehicles was studied via employing a sliding mode controller with neural networks \cite{27}. By mean of the graph theory, a distributed consensus formation control protocol was proposed \cite{17}. Researchers developed a formation control policy based on a self-organizing map neural network to obtain the coordinated behavior \cite{18}.} Nevertheless, the foregoing methods either addressed the 2-dimensional simplified scenarios, or employed the kinematic or reduced dynamic model for formation control synthesis. Considering higher dimensional and more complicated dynamics, the simulation study on formation control for multiple marine vehicles with full dynamics was addressed, and the gravity uncertainty was also taken into account \cite{5}. To alleviate the effects caused by data transmission delays and packet dropouts and reduce the mutual information exchange among the vehicles, line-of-sight measurement-based leader-follower formation control was studied, and the backstepping controller was then developed using Lyapunov's method \cite{24}. The receding horizontal formation tracking control constrained with underactuated dynamics was investigated, and the corresponding receding horizon controller was presented \cite{20}. Formation control for multiple underactuated underwater vehicles was addressed based on the small gain method which allows to design decentralized controllers with stability guarantees \cite{26}. In order to achieve finite-time convergence and meanwile adapt to the changeable dynamic parameters, a robust adaptive fast terminal sliding mode formation controller was proposed for multi-underactuated AUV systems \cite{XIA2021108903}.

 Almost all of aforementioned research utilized the leader-following coordinating structure to design the formation controller for each vehicle. As we mentioned earlier, this type of approach was highly dependent on the performance of the leading vehicle and was not able to obtain a good coordination among the neighboring vehicles. As such, in order to further improve the coordination performance, it is quite necessary to incorporate the information of neighboring vehicles when synthesizing the control law. Different from the existing work, this paper is concerned with consensus formation tracking control in 3-dimensional space, where the vehicles' formation is controlled not only relying on the information from leader, but also their neighbors, and meanwhile it is also assumed that each vehicle may suffer from the unknown disturbances either because of hydrodynamic-related phenomena or time-varying marine environments. On the other hand, it is well known that the sliding mode control (SMC) owing to its robustness and simplicity has been attracted much attention in the area of robotics and networked systems, as shown in above literature. The readers are recommended to refer to the survey paper \cite{SMC} for more detailed review on potential applications of SMC schemes in networked systems. As a result, the goal of this paper is to develop a distributed sliding mode control protocol for multi-AUV systems to achieve a good formation tracking performance, and particularly taking into consideration that the conventional SMC scheme is always suffering from the chattering issue as well as the vulnerability to the noises, a novel bioinspired handling method is employed. The main contributions of this paper are listed as follows:
\begin{enumerate}
\item{Distributed formation tracking control of multi-AUVs subject to the hydrodynamic parameter uncertainties and unknown exogenous disturbances in 3-dimensional space is addressed, and particularly full dynamics of the AUVs is adopted.}
\item{Consensus problem is accounted for in the design of distributed formation controller to improve the coordination performance of the overall formation system.}
\item{A novel distributed bioinspired sliding mode control protocol is developed to achieve the considered control objectives, where the control law is realized in a fully distributed manner. Meanwhile, the chattering phenomena and non-smoothness arising in the conventional SMC scheme are completely eliminated, rendering a more practical control activity, and the robustness to the noises is also enhanced.}
\item{Rigorous stability analysis is provided based on the Lyapunov theory to theoretically guarantee the anticipated performance of proposed distributed formation control protocol.}

\end{enumerate}

The rest of the paper is arranged as follows. The problem formulation and preliminary are presented in the next section. In Section \ref{s3}, bioinspired sliding mode controller is derived, and the stability analysis of resulting system is discussed. The simulation results are shown in Section \ref{s4}. Section \ref{s5} draws the conclusion.

% \section{The Design, Intent, and \\ Limitations of the Templates}
% The templates are intended to {\bf{approximate the final look and page length of the articles/papers}}. {\bf{They are NOT intended to be the final produced work that is displayed in print or on IEEEXplore\textsuperscript{\textregistered}}}. They will help to give the authors an approximation of the number of pages that will be in the final version. The structure of the \LaTeX\ files, as designed, enable easy conversion to XML for the composition systems used by the IEEE. The XML files are used to produce the final print/IEEEXplore pdf and then converted to HTML for IEEEXplore.

% \section{Where to Get \LaTeX \ Help --- User Groups}
% The following online groups are helpful to beginning and experienced \LaTeX\ users. A search through their archives can provide many answers to common questions.
% \begin{list}{}{}
% \item{\url{http://www.latex-community.org/}} 
% \item{\url{https://tex.stackexchange.com/} }
% \end{list}

\section{PRELIMINARY AND PROBLEM FORMULATION}\label{s2}
In this section, the mathematical model of AUVs in three-dimensional space is presented, and the objective of formation control for multi-AUV system is described. Moreover, the preliminary knowledge on graph theory is also presented. 

\subsection{Preliminary on graph theory}\label{s2.1}
For the considered group of underwater vehicles, it is assumed that each vehicle can be viewed as a node, and the communication topology associated with the networked system among multiple AUVs can then be represented by a weighted directed graph $G = \{ {V,E,A} \} $. As for a simple time-invariant graph $G$, it is composed of the vertex set $V = \{ { \nu _1, \nu _2, \ldots, \nu _N } \}$ , the edge set $E \subseteq V \times V$, and the weighted adjacency matrix $A = \left[ {{a_{ij}}} \right] \in \mathbb{R}^{\text{N} \times \text{N}}$. The element $\nu _i$ in vertex set $V$ denotes vehicle $i$, and the index $i$ belongs to a finite index set $\Gamma  = \left\{ {1, \ldots ,N} \right\}$. If there exists the information exchange between AUV $i$ and AUV $j$, then, say, there is a edge between AUVs $i$ and $j$, i.e., $\left( {{\nu _i},{\nu _j}} \right) \in E$, and ${a_{ij}} = {a_{ji}} > 0$. Particularly, call vehicle $j$ a neighbor of vehicle $i$, and the set of neighbors is represented by ${N_i} = \left\{ {j | {\left( {{\nu _i},{\nu _j}} \right) \in E} } \right\}$. Otherwise, there is no edge among them, and $a_{ij} = a_{ji} = 0$. Moreover, we define $a_{ii} = 0$ for all $i \in \Gamma$, and the out-degree $d_{i} = \sum _{j \in N_i} {a_{ij}}$ associated with the node $i$. Afterwards, the degree matrix together with the Laplacian matrix of the graph $G$ can then be defined as $D = \text{diag} \left\{ {d_1, \ldots, d_N}  \right\} \in \mathbb{R} ^{\text{N} \times \text{N}}$ and $L = D-A$, respectively. A path in graph is a sequence that is consisted of a set of successive adjacent nodes, starting from node $i$ and ending at node $j$. If any two nodes in a graph $G$ have at least one path, then, say, graph $G$ is connected.

In order to make the group of AUVs move along with the desired path as a whole, a reference trajectory must be defined ahead of time. The availability to the information of reference trajectory for $i$-th AUV is indicated by a parameter $b_i$; that is, if AUV $i$ is able to access this information,then $b_i >0$; $b_i = 0$, otherwise. Let $B = \text{diag} \left\{ {b_1, \ldots, b_N} \right\}$. 
\newtheorem{assumption}{Assumption}
\begin{assumption} \label{assumption1}
For the considered multi-AUV formation control network, graph $G$ is connected, and moreover there is at least one AUV able to receive the information of reference trajectory, i.e., the elements of $B$ are not all equal to zero.
\end{assumption}

\newtheorem{lemma}{Lemma}
\begin{lemma}\label{lemma1}
if Assumption \ref{assumption1} holds, then matrix $L+B$ is positive definite.
\end{lemma}

\begin{figure}[!t]
\centering
\includegraphics[width=2.5in]{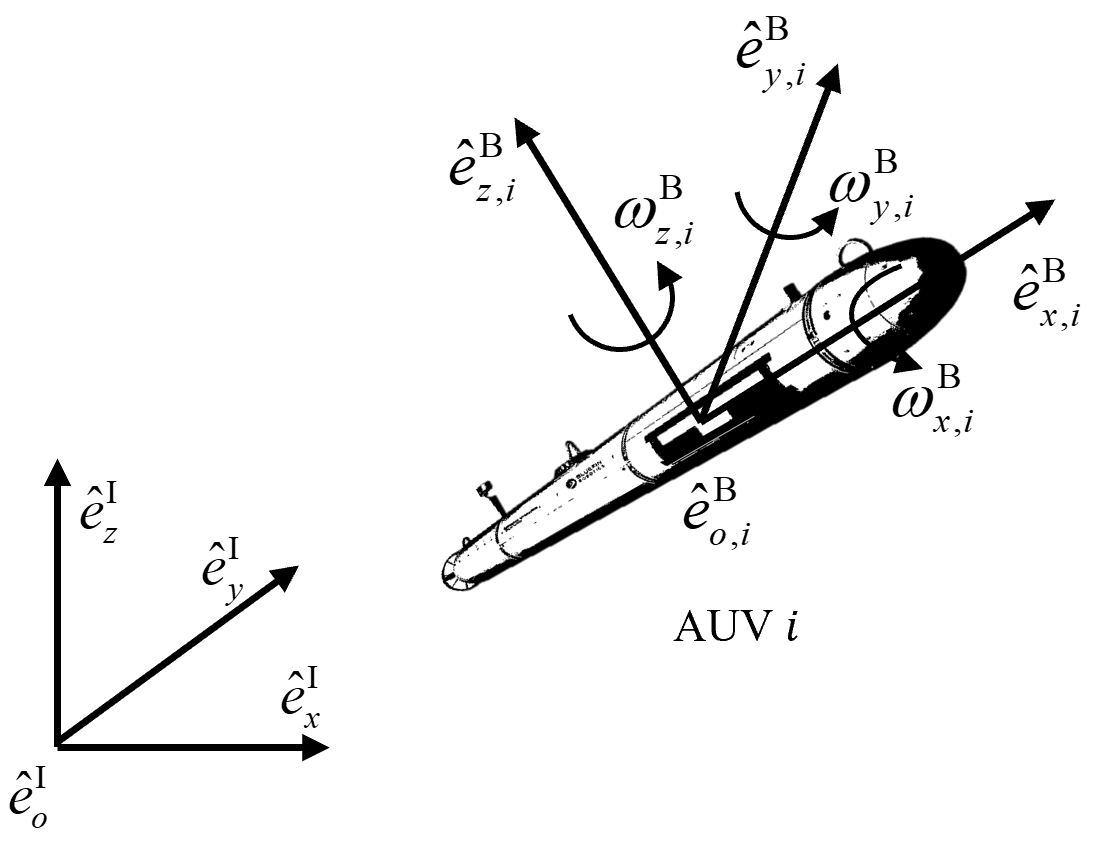}
\caption{Schematic diagram of $i$-th AUV .}
\label{fig1}
\end{figure}

\subsection{Problem formulation}
The formation control of $N$ numbers of autonomous underwater vehicles is investigated in this article. As presented in the work of Yang \textit{et al.} \cite{yang2013brief}, the dynamics of $i$-th AUV $(i \in \Gamma)$ can be modeled as
\begin{equation}\label{eq1}
\begin{split}
& \dot  \eta _i = J_i (\eta_{2,i})v_i, \\
& M_i \dot v_i + C_i (v_i) v_i + D_i (v_i)v_i = \tau _i + d_i ,
\end{split}
\end{equation}
where $\eta _i = \left[\eta _{1,i} \eta _{2,i} \right] ^{\rm T} \in \mathbb{R}^{6 \times 1}$, $\eta _{1,i} = \left[x_i, y_i, z_i \right] ^{\rm T} \in \mathbb{R}^{3 \times 1}$,  $\eta _{2,i} = \left[\phi_i, \theta_i, \psi_i \right] ^{\rm T} \in \mathbb{R}^{3 \times 1}$ denote the position and orientation of $i$-th AUV, respectively, which are expressed in the Earth-fixed frame $\hat E^{\rm I} = \left\{ {\hat e_o^{\rm I}, \hat e_x^{\rm I}, \hat e_y^{\rm I},\hat e_z^{\rm I}} \right\}$, and $ v_i = \left[v_{1,i}, v_{2,i} \right] ^{\rm T} \in \mathbb{R}^{6 \times 1}$, $v_{1,i} = \left[v_{x,i}, v_{y,i}, v_{z,i} \right] ^{\rm T} \in \mathbb{R}^{3 \times 1}$, $v_{2,i} = \left[\omega_{x,i}, \omega_{y,i}, \omega_{z,i} \right] ^{\rm T} \in \mathbb{R}^{3 \times 1}$ are the $i$-th AUV’s translational and rotational velocities, respectively, which are described in vehicle’s body-fixed frame $\hat E^{\rm B} = \left\{ {\hat e_{o,i}^{\rm B}, \hat e_{x,i}^{\rm B}, \hat e_{y,i}^{\rm B},\hat e_{z,i}^{\rm B}} \right\}$. The schematic of $i$-th AUV is illustrated in Fig. \ref{fig1}. {  $J_i (\eta_{2,i}) \in \mathbb{R}^{6 \times 6}$ denotes the coordinate-transform Jacobian matrix whose entries are time-varying.} $M_i \in \mathbb{R}^{6 \times 6}$ is the inertia matrix; $C_i(v_i) \in \mathbb{R}^{6 \times 6}$ is the Coriolis and centripetal matrix, and $D_i(v_i) \in \mathbb{R}^{6 \times 6}$ is the hydrodynamic damping matrix. $\tau _i \in \mathbb{R}^{6 \times 1}$ denotes the generalized control input vector of $i$-th AUV, { and $d_i \in \mathbb{R}^{6 \times 1}$ is the lumped disturbance, describing both the hydrodynamic parameter uncertainties as well as the maritime disturbances induced by the wind, waves and ocean currents. Jacobian matrix is defined as $J_i (\eta_{2,i}) = \text{diag} \left\{ {J_{1,i} (\eta_{2,i}), J_{2,i} (\eta_{2,i})} \right\}$, and it satisfies}
\begin{equation*}
    J_{1,i} (\eta_{2,i}) = \left[
    \begin{array}{cc}\cos{\theta _i} \cos{\psi _i} & \sin{\phi_i} \sin{\theta_i} \cos{\psi _i} - \cos{\phi_i} \sin{\psi_i} \\ 
    \cos{\theta _i} \sin{\psi _i}  & \cos{\phi_i} \cos{\psi _i} + \sin{\phi_i} \sin{\theta_i} \sin{\psi_i}  \\
    -\sin{\theta_i} & \sin{\phi_i} \cos{\theta_i} \end{array} \right.
\end{equation*}
\begin{equation*}
     \left.
    \begin{array}{c} \sin{\phi_i} \sin{\psi _i} + \cos{\phi_i} \sin{\theta_i} \cos{\psi_i} \\
    \cos{\phi_i} \sin{\theta_i} \sin{\psi _i} + \cos{\phi_i} \sin{\psi_i}\\
     \cos{\phi_i} \cos{\theta_i}   \end{array} \right],
\end{equation*}
\begin{equation*}
    J_{2,i} (\eta_{2,i}) = \begin{bmatrix}
    1 & \sin{\phi_i} \tan{\theta_i} & \cos{\phi_i} \tan{\theta_i}\\
    0 & \cos{\phi_i} & -\sin{\phi_i} \\
    0 & \sin{\phi_i} / \cos{\theta_i} & \cos{\phi_i} / \cos{\theta_i}
    \end{bmatrix}.
\end{equation*}

In order to circumvent the singularity of the Jacobian matrix, it is required to guarantee $\left| {\theta _i} \right| \ne \pi /2$. {The vessels' inertia matrix $M_i$ with added mass and the hydrodynamic damping matrix $ D_i$ are both diagonal and satisfy }
\begin{align*}
    M_i &= \text{diag} \left\{ { m_i - \beta_{\dot v x,i}, m_i - \beta_{\dot v y,i}, m_i - \beta_{\dot v z,i}, I_{x,i}, I_{y,i}, I_{z,i} } \right\},\\
    D_i &= -\text{diag} \left\{ { \beta_{ v x,i}, \beta_{ v y,i}, \beta_{v z,i},  \beta_{ \omega x,i}, \beta_{ \omega y,i}, \beta_{\omega z,i} } \right\},
\end{align*}
where $m_i$ and $I_{x,i}$, $I_{y,i}$, $I_{z,i}$ are the mass and rotational inertia of the $i$-th AUV, respectively; $\beta_{\dot v x,i}$, $\beta_{\dot v y,i}$, $\beta_{\dot v z,i}$, $\beta_{ v x,i}$, $\beta_{ v y,i}$, $\beta_{ v z,i}$, $\beta_{ \omega x,i}$, $\beta_{ \omega y,i}$ and $\beta_{ \omega z,i}$ are the hydrodynamic parameters. The matrix $C_i(v_i)$ is made up of the Coriolis and centripetal matrix $C_{R,i}(v_i)$ as well as the hydrodynamic extra term $C_{A,i}(v_i)$, defined as
\begin{align*}
    C_i(v_i) &= C_{R,i}(v_i) + C_{A, i}(v_i), \\
    & = \begin{bmatrix}
    \mathbf{0}_{3 \times 3} & C_{R1,i}(v_i) \\
    C_{R1,i}(v_i) & C_{R2,i}(v_i)
    \end{bmatrix} + \begin{bmatrix}
     \mathbf{0}_{3 \times 3} & C_{A1,i}(v_i) \\
    C_{A1,i}(v_i) & C_{A2,i}(v_i)
    \end{bmatrix},
\end{align*}
 where
 \begin{align*}
     C_{R1,i}(v_i) & = \begin{bmatrix}
     0 & m_i v_{z,i} & - m_i v_{y,i} \\
     -m_i v_{z,i} & 0 & m_i v_{x,i} \\
     m_i v_{y,i} & - m_i v_{x,i} & 0
     \end{bmatrix},\\
     C_{R2,i}(v_i) & = \begin{bmatrix}
     0 & I_{z,i} \omega _{z.i} & - I_{y,i} \omega{y,i} \\
     -I_{z,i} \omega _{z.i} & 0 & I_{x,i} \omega{x, i} \\
      I_{y,i} \omega{y,i}  & - I_{x,i} \omega{x, i} & 0
     \end{bmatrix}, \\
     C_{A1,i}(v_i) & = \begin{bmatrix}
     0 & \beta _{\dot v z, i} v_{z,i} & - \beta _{\dot v y, i} v_{y,i} \\
     - \beta _{\dot v z, i} v_{z,i} & 0 & \beta _{\dot v x, i} v_{x,i} \\
     \beta _{\dot v y, i} v_{y,i} & -\beta _{\dot v x, i} v_{x,i} & 0
     \end{bmatrix}, \\
    C_{A2,i}(v_i) & = \begin{bmatrix}
     0 & - \beta _{\dot \omega z, i} \omega_{z,i} & \beta _{\dot \omega y, i} v_{y,i} \\
     \beta _{\dot \omega z, i} \omega _{z,i} & 0 &  - \beta _{\dot \omega x, i} \omega_{x,i} \\
      - \beta _{\dot \omega y, i} \omega _{y,i} & \beta _{\dot \omega x, i} \omega_{x,i} & 0
     \end{bmatrix}, 
\end{align*}
and ${\beta _{\dot \omega x,i}}$, ${\beta _{\dot \omega y,i}}$, ${\beta _{\dot \omega z,i}}$ are the hydrodynamic parameters.

\newtheorem{remark}{Remark}
\begin{remark}\label{remark1}
It is worthwhile noting that the dynamics of AUV is a typical nonlinear system as described in \eqref{eq1}, having 6 degree of freedom (DOF), i.e., 3 translational DOF and 3 rotational DOF, and subject to the external disturbances as well as numerous uncertain hydrodynamic related parameters.
\end{remark}

In formation control problem, a desired formation pattern of group of AUVs is formed uniquely through the specific relative postures, i.e., positions and orientations, between the vehicles $i$ and $j$, $(i$, $j \in \Gamma)$; let the desired postures  $\Delta _{ij} = \left[ \delta _{x,ij}, \delta _{y,ij}, \delta _{z,ij}, \delta _{\phi,ij}, \delta _{\theta,ij}, \delta _{\psi,ij} \right] ^{\rm T} \in  \mathbb{R}^{6 \times 1}$. In most practical applications, the orientation of each AUV is required to be aligned; that is, the relative attitudes $\left[ \delta _{\phi,ij}, \delta _{\theta,ij}, \delta _{\psi,ij} \right] ^{\rm T} $ of AUVs are set to $\mathbf{0}_{3\times 1}$. In addition to the formation maintenance, a team of vehicles are also employed to track a desired trajectory in many maneuver missions. Let $\eta _1^d = \left[ x^d, y^d, z^d \right]^{\rm T} \in  \mathbb{R}^{3 \times 1} $ be the prescribed differentiable trajectory of the formation center, which can be viewed as a virtual leader in the group. { Based on the defined formation pattern, i.e., $\Delta _{ij}$, each vehicle’s desired trajectory $\eta _{1,i}^d$ associated with the virtual leader can then be derived accordingly.} $\eta _2^d = \left[ \phi^d, \theta^d, \psi^d \right]^{\rm T} \in  \mathbb{R}^{3 \times 1}$ denotes the desired differentiable attitude of the virtual leader, and combining position and orientation, let $\eta ^d = \left[ \eta _1^d, \eta _2^d \right] ^{\rm T}$ be the desired posture of virtual leader. {The objective of this paper is aimed to design a distributed formation control law for each individual in the group with full nonlinear dynamics and affected by the disturbances which are described in \eqref{eq1} so that the coordinated motion among AUVs can be achieved; in particular, the following control requirements are satisfied: 1) The AUVs can maintain a desired formation shape. 2) The group of AUVs can follow up a predefined trajectory as a whole.}

\section{FORMATION CONTROL PROTOCOL DESIGN}\label{s3}
{In this section, systematic design procedures for achieving solution for consensus formation control are presented.} First, the distributed sliding model controller is designed. Next, to eliminate the impacts of high-frequency oscillation that is caused by the employment of sign function, a bioinspired neurodynamics is introduced to avoid the use of nonlinear operation in control design. The main advantage of proposed controller is that it can provide smoother control inputs, which is much more significant for the actuators to generate realistic forces or moments. At the same time, a good performance in terms of disturbance rejection is still preserved. {Finally, the input-to-state stability of overall closed-loop system is proved.}

\subsection{Sliding mode control design}
The consensus cooperative tracking errors for $i$-th AUV $(i \in \Gamma)$ are defined as 
\begin{equation}\label{eq2}
\begin{split}
    e_{1,i} &= \sum\limits_{j \in N_i} {a_{ij} (\eta _i- \eta _j -\Delta _{ij}) + b_i (\eta _i -\eta _i ^d)},\\
    \dot e_{1,i} &= \sum\limits_{j \in N_i} {a_{ij} ( \dot \eta _i- \dot \eta _j) + b_i (\dot \eta _i - \dot \eta _i ^d)}.
\end{split}
\end{equation}
Here, the non-negative constant $a_{ij}$ indicates the information exchange between $i$-th vehicle and its neighbors $j \in N_i$, and another non-negative constant $b_i$ denotes whether or not $i$-th vehicle is able to access the information of the desired reference trajectory. $\Delta _{ij}$ denotes the relative posture (position and orientation) between $i$-th AUV and its neighbors, which determines the desired formation pattern needed to maintain. Note that since the whole group is intended to track the common reference trajectory, the desired velocities and accelerations of AUVs are the same; that is, $\dot\eta _i^d = \dot\eta ^d$, and $\ddot\eta _i^d = \ddot\eta ^d$, $(i \in \Gamma)$, where $ \dot\eta ^d$ and $ \ddot\eta ^d$ represent the desired velocities and accelerations of the whole group, respectively.

Taking the time derivative of $\dot e_{1,i}$, the following error dynamics for $i$-th AUV can be obtained
\begin{equation}\label{eq3}
    {\ddot e_{1,i}} = \sum\limits_{j \in {N_i}} {{a_{ij}}\left( {{{\ddot \eta }_i} - {{\ddot \eta }_j}} \right) + {b_i}\left( {{{\ddot \eta }_i} - \ddot \eta ^d} \right)},
\end{equation}
where $\ddot \eta _i$ and $\ddot \eta _j$ represent the dynamics of $i$-th AUV and its neighbors $j \in {N_i}$, respectively. According to the $i$ AUV’s kinematic and dynamic model as described in \eqref{eq1}, the expression of $\ddot \eta _i$ can be derived as follows
\begin{equation}\label{eq4}
    {\ddot \eta _i} = {A_i}\left( {{v_i},{\eta _i}} \right){v_i} + {J_i}\left( {{\eta _i}} \right){B_i}{\tau _i} + {J_i}\left( {{\eta _i}} \right){B_i}{d_i},
\end{equation}
where 
\begin{equation*}
    {A_i}\left( {{v_i},{\eta _i}} \right) = {\dot J_i}\left( {{\eta _i}} \right) - {J_i}\left( {{\eta _i}} \right){B_i}{C_i}\left( {{v_i}} \right) - {J_i}\left( {{\eta _i}} \right){B_i}{D_i}\left( {{v_i}} \right),
\end{equation*}
and  ${B_i} = {M_i}^{ - 1}$.

\begin{assumption} \label{assumption2}
The disturbance term ${J_i}\left( {{\eta _i}} \right){B_i}{d_i}$ can be bounded as 
\begin{equation}\label{eq5}
    \left\| {{J_i}\left( {{\eta _i}} \right){B_i}{d_i}} \right\| \le {\tilde \lambda _i}, \quad i \in \Gamma, 
\end{equation}
where $\tilde \lambda _i$ is a known positive constant. {This parameter typically depends on the marine situations.}
\end{assumption}

Letting 
\begin{equation}\label{eq6}
\begin{split}
    {{\bar e}_1} & = {\left[ {{e_{1,1}},{e_{1,2}}, \ldots ,{e_{1,N}}} \right]^{\rm T}},\\
    {{\bar e}_2} &= {\left[ {{{\dot e}_{1,1}},{{\dot e}_{1,2}}, \ldots ,{{\dot e}_{1,N}}} \right]^{\rm T}},
\end{split}
\end{equation}
and 
\begin{equation}\label{eq7}
    \ddot \eta  = {\left[ {{{\ddot \eta }_1},{{\ddot \eta }_2}, \ldots ,{{\ddot \eta }_N}} \right]^{\rm T}},
\end{equation}
the global error dynamics of overall system can be written as
\begin{equation}\label{eq8}
\begin{split}
    {{\dot{\bar e}}_1} &= {{\bar e}_2},\\
    {{\dot {\bar e}}_2} &= \left( {L + B} \right)\left( {\ddot \eta  - {{\mathbf{1}}_N}{{\ddot \eta }^d}} \right),
\end{split}
\end{equation}
where matrices $L$ and $B$ are defined in \ref{s2.1}, which describes the communication topology of entire system. ${{\mathbf{1}}_N}$ represents a $N$-dimensional vector whose entries are all equal to $1$.

The consensus formation control objective, now, is cast into designing a distributed control law so that the global error dynamics \eqref{eq8} can be stabilized at the origin. Next, sliding mode control design is employed and can be divided into the following two steps.

\textit{Step1: Define the sliding mode surface}

Let 
\begin{equation}\label{eq9}
    s = {\left[ {{s_1},{s_2}, \ldots {s_N}} \right]^{\rm T}},
\end{equation}
and the sliding mode variable $s$ is defined as
\begin{equation}\label{eq10}
    s = {k_1}\left( {L + B} \right){\bar e_1} + {\bar e_2},
\end{equation}
where $k_1$ is a positive constant needed to be designed. As we can see, so long as the sliding mode variable $s$ is able to reach to $\mathbf{0}$ in finite time, i.e., $s\left( t \right) \equiv 0$, $t\ge T$, say, system reached the sliding mode surface, the behavior of $\bar e_1$ then will be solely governed by the sliding mode surface $ s = \dot {\bar{ e}}_1 + {k_1} \left( {L + B} \right) {\bar e_1} = 0$ in all future time, which leads to $\bar e\left( t \right) \to 0$, as $t \to \infty $. { In other words, the system's behaviour will be firmly restricted into the surface and sliding on it until $\bar e\left( t \right) \to 0$. Most importantly, as we observed there is no disturbance terms enforced on the sliding mode surface while the original system dynamics may be subject to all kinds of disturbances, and thus robustness properties can be achieved once system enters into the sliding mode surface.  At next step, our goal is to design sliding mode control law to render system reach to the sliding mode surface and stay on it in all finite time.}

\textit{Step2: Design sliding mode control law}

In this step, it is shown that the sliding mode surface can be reached in finite time, and the system will stay at the surface for all future time.

Choose the following Lyapunov function candidate
\begin{equation}\label{eq11}
    {V_s} = \frac{1}{2}{s^{\rm T}}{\left( {L + B} \right)^{ - 1}}s,
\end{equation}
whose time derivative is calculated as
\begin{equation}\label{eq12}
    {\dot V_s} = {s^{\rm T}}{\left( {L + B} \right)^{ - 1}}\dot s.
\end{equation}
Based on the definition of sliding mode variable $s$, as shown in \eqref{eq10}, $ \dot s$ can be obtained
\begin{equation}\label{eq13}
    \dot s = {k_1}(L + B){\bar e_2} + {\dot{{\bar e}}_2}.
\end{equation}
Due to the global error dynamics \eqref{eq8}, yield
\begin{equation}\label{eq14}
    \dot s = {k_1}(L + B){\bar e_2} + \left( {L + B} \right)\left( {\ddot \eta  - {{\mathbf{1}}_N}{{\ddot \eta }^d}} \right).
\end{equation}
{By substituting \eqref{eq14} into \eqref{eq12}, we have}
\begin{equation}\label{eq15}
    {\dot V_s} = {s^{\rm T}}{\left( {L + B} \right)^{ - 1}}\left[ {{k_1}\left( {L + B} \right){{\bar e}_2} + \left( {L + B} \right)\left( {\ddot \eta  - {{\mathbf{1}}_N}{{\ddot \eta }^d}} \right)} \right].
\end{equation}

Let
\begin{align*}
     & \bar A\left( {v,\eta } \right) = \text{diag}\left\{ {{A_1}\left( {{v_1},{\eta _1}} \right), \ldots ,{A_1}\left( {{v_N},{\eta _N}} \right)} \right\},\\
     &\bar B = \text{diag} \left\{ {{B_1}, \ldots ,{B_N}} \right\},\\
     &\bar J\left( \eta  \right) =  \text{diag}\left\{ {{J_1}\left( {{\eta _1}} \right), \ldots ,{J_N}\left( {{\eta _N}} \right)} \right\},\\
     & \tau  = {\left[ {{\tau _1},{\tau _2}, \ldots {\tau _N}} \right]^{\rm T}},
\end{align*}
and propose the following distributed control law
\begin{equation}\label{eq16}
    \tau  = {\left[ {\bar J\left( \eta  \right)\bar B} \right]^{ - 1}}\left[ { - \bar A\left( {v,\eta } \right)v + \tau '} \right],
\end{equation}
where $\tau ' $ is an auxiliary control variable and is designed as
\begin{equation}\label{eq17}
    \tau ' =  - {k_1}{\bar e_2} + {{\bf{1}}_N}{\ddot \eta ^d} - {\beta _0} \text{sign} \left( s \right),
\end{equation}
where $\beta _0$ is a positive designable constant whose value is dependent on the upper bound of disturbances and is needed to satisfy with ${\beta _0} \ge {\sup _{i \in \Gamma }}\left\{ {{{\tilde \lambda }_i}} \right\} + \varepsilon $, where $\varepsilon$ is an arbitrary positive constant.

Combing with AUV's dynamics \eqref{eq4} as well as \eqref{eq7} and plugging \eqref{eq16} into the \eqref{eq15} result in
\begin{equation}\label{eq18}
    \dot V _s = {s^{\rm T}}\left[ {{k_1}{{\bar e}_2} - {{\bf{1}}_N}{{\ddot \eta }^d} + \tau ' + \bar J\left( \eta  \right)\bar Bd} \right],
\end{equation}
where $d = {\left[ {{d_1}, \ldots {d_N}} \right]^{\rm T}}$. Substituting the auxiliary control variable \eqref{eq17} into \eqref{eq18}, we may end up with
\begin{equation}\label{eq19}
\begin{split}
{{\dot V}_s} &= {s^{\rm T}}\left[ { - {\beta _0}sign\left( s \right) + \bar J\left( \eta  \right)\bar Bd} \right]\\
 &\le  - \sum\limits_{i = 1}^N {({\beta _0} - \left\| {{J_i}\left( {{\eta _i}} \right){B_i}{d_i}} \right\|)\left| {{s_i}} \right|} \\
 &\le  - \sum\limits_{i = 1}^N \varepsilon  \left| {{s_i}} \right| \le  - \sqrt {\frac{2}{{{\lambda _{\max }}{{\left( {L + B} \right)}^{ - 1}}}}} {V^{1/2}}.
\end{split}
\end{equation}
By taking integral on both sides, we can conclude
\begin{equation}\label{eq20}
    \sqrt {{V_s}\left( t \right)}  \le  - \sqrt {\frac{1}{{2{\lambda _{\max }}{{\left( {L + B} \right)}^{ - 1}}}}} t + {V_s}\left( 0 \right).
\end{equation}
Above inequality shows that under proposed sliding mode control law \eqref{eq16} and \eqref{eq17} the sliding variable $s$ will reach to zero in finite time and stay there for all future time, i.e., $s \equiv 0$, $t \ge T$, where 
\begin{equation}\label{eq21}
    T = \sqrt {2{\lambda _{\max }}{{\left( {L + B} \right)}^{ - 1}}} {V_s}\left( 0 \right).
\end{equation}
Once system is sliding on the surface $s \equiv 0$, the overall behavior of the system is then determined by the linear differential equation
\begin{equation*}
    {\dot {\bar e}}_1 + {k_1}\left( {L + B} \right){\bar e_1} = 0.
\end{equation*}
Moreover, as the Lemma \ref{lemma1} shows $L + B$ is a positive definite matrix, it follows that ${\bar e_1} \to 0$ as $t \to \infty $. When ${\bar e_1} \to 0$, ${\dot {\bar e}}_1$ will also tend to $0$; then, according to \eqref{eq8}, ${\bar e_2} \to 0$ as well. {That is, the error dynamics \eqref{eq8} is asymptotically stable and its converging rate is dependent on the $k_1$.}

\subsection{Bioinspired control design}
Last section we presented the design process of sliding mode control for consensus formation control of multiple AUV systems, and the asymptotical stability is achieved even in the presence of unknown bounded disturbances, which actually relies on the fact that the nonlinear sign function is employed in SMC control law. Theoretically, this type of control is able to handle any matched bounded disturbances enforced on the system. However, the resulting high frequency non-smooth control activities can hardly be realized in reality  due to the physical limitations on actuators. Thus, the theoretical robust performance cannot always be obtained in applications; furthermore, the stability of overall system might be affected due to the excitation of high-frequency unmodeled dynamics. As a result, a neural dynamics-based control approach is proposed and studied to address this critical issue encountered in conventional sliding mode control.

{In this section a neurodynamic model, termed shunting model, is employed to take place of the classic nonlinear sign function from which the chattering originates in conventional SMC scheme so that the use of discontinuous control is avoided and chattering phenomenon is completely vanishing. In fact, we will show that due to the integration with the shunting model many favorable properties can be obtained in the resulting formation system.} 

Shunting model is one of the most popular bioinspired dynamic models that was first proposed in the work of \cite{OGMEN1990487}, which was built to try to characterize the adaptive behavior of membrane to the ever changing environment. Due to its desirable characteristics, in the past few decades it has been incorporated in various robotic systems to develop efficient algorithms for real-time path planning as well as feedback control \cite{yang2003real,yuan2003virtual,yang2011bioinspired,zhu2021bio, 9136718}. The shunting equation can be described as follows
\begin{equation}\label{eq22}
    {\dot x_i} =  - {a_i}{x_i} + \left( {{b_i} - {x_i}} \right)s_i^ +  - \left( {{d_i} + {x_i}} \right)s_i^ - ,
\end{equation}
where $ x_i$ represents the neural activity of  $i$-th neuron, associated with some nonnegative constant parameters $a_i$, $b_i$ and $d_i$ indicating the rate of response, the maximum and minimum values of the neuron activities, respectively. Signals $s_i^+$ and $s_i^-$ represent the effects of environmental changes, i.e., the corresponding excitatory and inhibitory inputs to the neuron, respectively. 

{
\begin{remark} \label{remark2}
It is interesting to note that the presented shunting equation describes a dynamic characteristic of a neuron, and thus a more consistent behavior of the neural activities can be obtained, while the environmental changes may exhibit a highly evident discontinuity or non-smoothness. In particular, the transient stage can be adjustable by the parameter $a_i$, and furthermore, due to the nonlinearity nature of the shunting model larger environmental changes may result in a faster transient process. Besides of that, the outputs of the shunting model, i.e., the neural activities, are able to be bounded upper by the parameter $b_i$ and lower by the parameter $d_i$. It will be shown, in what follows, that such properties of the shunting model can be applied to significantly improve the performance of the conventional SMC scheme.

\end{remark}}
In order to apply the shutting model to formation control design, let
\begin{equation}\label{eq23}
\begin{split}
 s_i^+ &= 
 \begin{cases}
 s_i, &\text{if} \  s_i \ge{0},\\
 {0}, & {\text{otherwise}}.
 \end{cases}\\
  s_i^- &= 
 \begin{cases}
 -s_i, &\text{if} \  s_i <{0},\\
 {0}, & {\text{otherwise}}.
 \end{cases}
\end{split}
\end{equation}

Define ${s^ + } = {\left[ {s_1^ + , \ldots s_N^ + } \right]^{\rm T}}$ and ${x_s} = {\left[ {{x_1}, \ldots {x_N}} \right]^{\rm T}}$, and the shunting model-based SMC control law can be proposed as follows
\begin{equation}\label{eq24}
\begin{split}
    &\tau  = {\left[ {\bar J\left( \eta  \right)\bar B} \right]^{ - 1}}\left[ { - \bar A\left( {v,\eta } \right)v + \tau '} \right],\\
    &\tau ' =  - {k_1}{{\bar e}_2} + {{\bf{1}}_N}{{\ddot \eta }^d} - {\beta _0}{x_s},\\
    &{{\dot x}_s} =  - {A_s}{x_s} + \left( {{B_s} - {x_s}} \right){s^ + } - \left( {{D_s} + {x_s}} \right){s^ - },\\
    &s = {k_1}\left( {L + B} \right){{\bar e}_1} + {{\bar e}_2}.
\end{split}
\end{equation}
Here, ${A_s} = \text{diag}\left\{ {{a_1}, \ldots ,{a_N}} \right\}$, ${B_s} = \text{diag} \left\{ {{b_1}, \ldots ,{b_N}} \right\}$ and ${D_s} = \text{diag} \left\{ {{d_1}, \ldots ,{d_N}} \right\}$. 

{
\begin{remark}\label{remark3}
It is observed that the output of the shunting model, i.e., $x_s$, is used in the auxiliary control law $\tau'$, instead of the original switching function as shown in \eqref{eq17}. As a result, due to the benefits of the dynamic characteristic of the shunting equation, as mentioned previously in Remark \ref{remark2}, a more consistent and smoother control activities can be rendered, and thus the chattering issue is avoided completely. Additionally, the output of the shunting model is also allowed to be bounded. Such observations demonstrate the significant improvements for the conventional SMC approach for real-world applications, that is, the proposed bioinspired SMC scheme can lead to a chattering-free behavior and yield more practical control actions for actual actuators to implement.
\end{remark}
}
{
\begin{remark}\label{remark}
It is worth noting that while the robust asymptotical stability of the SMC scheme along the sliding mode surface is no longer held,  owing to the employment of shunting model, the resulting control strategy can still allow to achieve a good robustness against the bounded disturbances as well as the noises, which can be seen clearly in the next section on stability analysis of the proposed formation system.
\end{remark}
}

\subsection{Stability analysis}
To analyze the stability of overall AUV formation system, plugging the proposed distributed controller \eqref{eq24} into the global error dynamics \eqref{eq8} together with \eqref{eq4}, \eqref{eq6} and \eqref{eq7}, the following closed-loop system then can be obtained
\begin{align} 
    &{\dot {\bar e}}_1 =  - {k_1}\left( {L + B} \right){\bar e_1} + s, \label{eq25}\\
    &\dot s =  - \left( {L + B} \right){\beta _0}{x_s} + \left( {L + B} \right)\bar J\left( \eta  \right)\bar Bd,\label{eq26}\\
    &{\dot x_s} =  - {A_s}{x_s} + \left( {{B_s} - {x_s}} \right){s^ + } - \left( {{D_s} + {x_s}} \right){s^ - }. \label{eq27} 
\end{align}

It can be found that the resulting closed-loop system is a coupled perturbed system with state vectors $\left( {{{\bar e}_1},\dot s,{{\dot x}_s}} \right)$ as well as disturbances $d$. To be more specific, subsystem \eqref{eq25} and \eqref{eq26} are cascaded while subsystem \eqref{eq26} and \eqref{eq27} are of interconnected architecture. Hence, based on this feature the proof may proceed with two steps. That is, first step shows subsystem \eqref{eq25} is input-to-state stable, regarding $s$ as the input; second, the interconnected subsystems \eqref{eq26} and \eqref{eq27} are also input-to-state stable with the impacts of disturbances. Finally, on the basis of such two results one may conclude that the overall system \eqref{eq25}-\eqref{eq27} is input-to-state stable.

\textit{Step1: Stability of subsystem \eqref{eq25}}

Due to the characteristic of linearity, the solution of subsystem \eqref{eq25} can be readily written as 
\begin{equation}\label{eq28}
    {\bar e_1}\left( t \right) = {e^{ - {k_1}\left( {L + B} \right)t}}{\bar e_1}\left( 0 \right) + \int_0^t {{e^{ - {k_1}\left( {L + B} \right)\left( {t - \tau } \right)}}sd\tau } .
\end{equation}
It follows that
\begin{equation}\label{eq29}
\begin{split}
    \left\| {{{\bar e}_1}\left( t \right)} \right\| &\le {e^{ - \tilde kt}}\left\| {{{\bar e}_1}\left( 0 \right)} \right\| + \int_0^t {{e^{ - {k_1}\left( {L + B} \right)\left( {t - \tau } \right)}}\left\| {s\left( \tau  \right)} \right\|d\tau } \\
    &\le {e^{ - \tilde kt}}\left\| {{{\bar e}_1}\left( 0 \right)} \right\| + \frac{1}{{\tilde k}}\mathop {\sup }\limits_{0 \le \tau  \le t} \left\| {s\left( \tau  \right)} \right\|,
\end{split}
\end{equation}
{where $\tilde k = {\lambda _{\min }}\left( {{k_1}\left( {L + B} \right)} \right)$ . It can be concluded from \eqref{eq29} the ultimate bound of $\left\| {{{\bar e}_1}\left( t \right)} \right\|$ is proportional to the bound of $\left\| {s\left( t \right)} \right\|$, from which it is shown that the subsystem \eqref{eq25} is input-to-state stable; moreover, the larger $k_1$ is , the less ultimate bound will be.}

\textit{Step2: Stability of interconnected subsystems \eqref{eq26} and \eqref{eq27}}

Because of the employment of piecewise function $s^+$ and $s^-$ in subsystem \eqref{eq27}, the analysis shall proceed with two cases. First, when $s \ge 0$, the subsystem \eqref{eq27} can be rewritten as 
\begin{equation}\label{eq30}
    {\dot x_s} =  - \left( {{A_s} + S} \right) {x_s} + {B_s}s,
\end{equation}
where $S = \text{diag}\left \{ s_1, s_2, \ldots, s_N \right \}$. 

Let $\tilde x  = \left[ {s, x_s} \right]^{\rm T}$, and choose the following Lyapunov function candidate
\begin{equation} \label{eq31}
    {V_1} = \frac{1}{2} \tilde x ^{\rm T} Q \tilde x,
\end{equation}
where
\begin{equation}\label{eq32}
    Q = \begin{bmatrix}
    \frac{2}{\beta _0} {\left( {L+B} \right)^{-1}} & { -\left( {A_s + S} \right)}^{-1}\\
    { -\left( {A_s + S} \right)}^{-1} & {B_s^{-1}}
    \end{bmatrix},
\end{equation}
and the following condition is assumed to be satisfied
\begin{equation}\label{eq33}
    \frac{2}{{{\beta _0}}}{\left( {L + B} \right)^{ - 1}}B_s^{ - 1} - {\left( {{A_s}^{ - 1}} \right)^2} \succ 0,
\end{equation}
which guarantees $V_1$ to be positive definite. Then, taking the time derivative of $V_1$ along the subsystems \eqref{eq26} and \eqref{eq30} together with \eqref{eq32} yields
\begin{equation}\label{eq34}
\begin{split}
{{\dot V}_1} &= - {s^{\rm T}}{\left( {{A_s} + S} \right)^{ - 1}}s\\
& \quad - x_s^{\rm T}\left[ {B_s^{ - 1}\left( {{A_s} + S} \right) - {\beta _0}{{\left( {{A_s} + S} \right)}^{ - 1}}\left( {L + B} \right)} \right]{x_s}\\
& \quad + \frac{2}{{{\beta _0}}}{s^{\rm T}}\bar J\left( \eta  \right)\bar Bd - x_s^{\rm T}{\left( {{A_s} + S} \right)^{ - 1}}\left( {L + B} \right)\bar J\left( \eta  \right)\bar Bd\\
& \le - {s^{\rm T}}{\left( {{A_s} + S} \right)^{ - 1}}s\\
& \quad- x_s^{\rm T}\left[ {B_s^{ - 1}{A_s} - {\beta _0}{A_s}^{ - 1}\left( {L + B} \right)} \right]{x_s} + \frac{2}{{{\beta _0}}}{s^{\rm T}}\bar J\left( \eta  \right)\bar Bd\\
& \quad - x_s^{\rm T}{\left( {{A_s} + S} \right)^{ - 1}}\left( {L + B} \right)\bar J\left( \eta  \right)\bar Bd.
\end{split}
\end{equation}
Since $A_s$ and $S$ are both positive definite diagonal matrices, the first term in \eqref{eq34} is already negative definite with respect to $s$. As for the second term, it is assumed that the following matrix inequality holds
\begin{equation}\label{eq35}
    B_s^{ - 1}{A_s} - {\beta _0}{A_s}^{ - 1}\left( {L + B} \right) \succ 0.
\end{equation}
Define $P = B_s^{ - 1}{A_s} - {\beta _0}{A_s}^{ - 1}\left( {L + B} \right)$. The rest of the terms indicating the effects from disturbances can be addressed using the bounded approach. Applying the bounded disturbance condition \eqref{eq5}, the derivative of $V_1$ can be rewritten as 
\begin{equation}\label{eq36}
\begin{split}
    {{\dot V}_1} &\le  - {\zeta _1}{\left\| s \right\|^2} - {\zeta _2}{\left\| {{x_s}} \right\|^2} + \frac{{2 }}{{{\beta _0}}} {\left\| \tilde \lambda \right\|} \left\| s \right\| + {\zeta _3} {\left\| \tilde \lambda \right\|}\left\| {{x_s}} \right\|\\
    & =  - \left( {1 - \theta } \right){\zeta _1}{\left\| s \right\|^2} - \left( {1 - \vartheta } \right){\zeta _2}{\left\| {{x_s}} \right\|^2} \\
    & \quad- \theta {\zeta _1}{\left\| s \right\|^2} - \vartheta {\zeta _2}{\left\| {{x_s}} \right\|^2} + \frac{{2 }}{{{\beta _0}}} {\left\| \tilde \lambda \right\|}\left\| s \right\| + {\zeta _3} {\left\| \tilde \lambda \right\|} \left\| {{x_s}} \right\|\\
    & \le  - \left( {1 - \theta } \right){\zeta _1}{\left\| s \right\|^2} - \left( {1 - \vartheta } \right){\zeta _2}{\left\| {{x_s}} \right\|^2},\\
    & \qquad  \left( \forall \left\| s \right\| \ge \frac{{2 }}{{\theta {\beta _0}{\zeta _1}}}{\left\| \tilde \lambda \right\|}, \forall \left\| {{x_s}} \right\| \ge \frac{{{\zeta _3} }}{{\vartheta {\zeta _2}}}{\left\| \tilde \lambda \right\|} \right)
\end{split}
\end{equation}
{where $\tilde \lambda  = {\left[ {{{\tilde \lambda }_1}, \ldots, {{\tilde \lambda }_N}} \right]^{\rm T}}$, ${\zeta _1} = {\lambda _{\min }}\left( {{{\left( {{A_s} + S} \right)}^{ - 1}}} \right)$, ${\zeta _2} = {\lambda _{\min }}\left( {P} \right)$ and ${\zeta _3} = {\lambda _{\max }}\left( {{{\left( {{A_s} + S} \right)}^{ - 1}}\left( {L + B} \right)} \right)$. $\theta $ and $\vartheta$ are two numbers in the interval $\left( {0,1} \right)$.}

Equation \eqref{eq36} shows that ${\dot V_1} < 0$, when $\left\| s \right\| \ge \frac{{2\tilde \lambda }}{{\theta {\beta _0}{\zeta _1}}}$ and $\left\| {{x_s}} \right\| \ge \frac{{{\zeta _3}\tilde \lambda }}{{\vartheta {\zeta _2}}}$, from which it can be concluded that the interconnected subsystems \eqref{eq26} and \eqref{eq30} are input-to-state stable.

Likewise, when $s < 0$, the subsystem \eqref{eq27} can be reduced to
\begin{equation}\label{eq37}
    {\dot x_s} =  - \left( {{A_s} -S} \right){x_s} + {D_s}s.
\end{equation}
The analysis procedure is very similar to the previous case. First, choose Lyapunov function candidate as
\begin{equation} \label{eq38}
    {V_1} = \frac{1}{2} \tilde x ^{\rm T} Q' \tilde x,
\end{equation}
where
\begin{equation}\label{eq39}
    Q' = \begin{bmatrix}
    \frac{2}{\beta _0} {\left( {L+B} \right)^{-1}} & { -\left( {A_s - S} \right)}^{-1}\\
    { -\left( {A_s - S} \right)}^{-1} & {D_s^{-1}}
    \end{bmatrix},
\end{equation}
and assume the following condition is satisfied
\begin{equation}\label{eq40}
    \frac{2}{{{\beta _0}}}{\left( {L + B} \right)^{ - 1}}D_s^{ - 1} - {\left( {{A_s}^{ - 1}} \right)^2} \succ 0,
\end{equation}
which renders $V_2$ to be positive definite. Then, the derivative of $V_2$ along the subsystems \eqref{eq26} and \eqref{eq37} can be obtained
\begin{equation}\label{eq41}
\begin{split}
    {{\dot V}_2}  &\le  - {s^{\rm T}}{\left( {{A_s} -S} \right)^{ - 1}}s \\
    & \quad - x_s^{\rm T}\left[ {D_s^{ - 1}{A_s} - {\beta _0}{A_s}^{ - 1}\left( {L + B} \right)} \right]{x_s}\\
    & \quad + \frac{2}{{{\beta _0}}}{s^{\rm T}}\bar J\left( \eta  \right)\bar Bd\\
    & \quad - x_s^{\rm T}{\left( {{A_s} - S} \right)^{ - 1}}\left( {L + B} \right)\bar J\left( \eta  \right)\bar Bd.
\end{split}
\end{equation}
The term $A_s -S$ is already positive definite due to the fact that $s<0$, and we also need the following matrix inequality to be satisfied
\begin{equation}\label{eq42}
D_s^{ - 1}{A_s} - {\beta _0}{A_s}^{ - 1}\left( {L + B} \right) \succ 0.
\end{equation}
Define $P' = D_s^{ - 1}{A_s} - {\beta _0}{A_s}^{ - 1}\left( {L + B} \right)$. Similarly, as above ${\dot V_2}$ can be rewritten as 
\begin{equation}\label{eq43}
\begin{split}
    {\dot V_2} \le  - \left( {1 - \theta } \right){\zeta '_1}{\left\| s \right\|^2} - \left( {1 - \vartheta } \right){\zeta '_2}{\left\| {{x_s}} \right\|^2}, \\
    \left(\forall \left\| s \right\| \ge \frac{{2\tilde \lambda }}{{\theta {\beta _0}{{\zeta '}_1}}},\forall \left\| {{x_s}} \right\| \ge \frac{{{{\zeta '}_3}\tilde \lambda }}{{\vartheta {{\zeta '}_2}}} \right)
\end{split}
\end{equation}
where ${\zeta '_1} = {\lambda _{\min }}\left( {{{\left( {{A_s} - S} \right)}^{ - 1}}} \right)$, ${\zeta '_2} = {\lambda _{\min }}\left( {P'} \right)$ and ${\zeta '_3} = {\lambda _{\max }}\left( {{{\left( {{A_s} - S} \right)}^{ - 1}}\left( {L + B} \right)} \right)$. $\theta$ and $\vartheta$ are two numbers in the interval $\left( {0,1} \right)$.

It follows from \eqref{eq43} that ${\dot V_2} < 0$, when $\left\| s \right\| \ge \frac{{2\tilde \lambda }}{{\theta {\beta _0}{{\zeta '}_1}}}$ and $\left\| {{x_s}} \right\| \ge \frac{{{{\zeta '}_3}\tilde \lambda }}{{\vartheta {{\zeta '}_2}}}$, which shows that the interconnected subsystems \eqref{eq26} and \eqref{eq37} are input-to-state stable.

Since the subsystem \eqref{eq25} and interconnected subsystems \eqref{eq26} and \eqref{eq27} are all input-to-state stable, so long as the controller parameters are properly chosen such that the conditions \eqref{eq33}, \eqref{eq35}, \eqref{eq40} as well as \eqref{eq42} are satisfied, it is readily shown that the overall closed-loop system \eqref{eq25}-\eqref{eq27} is input-to-state stable. This completes the proof.

\begin{remark}\label{remark4}
It should be noted that due to the introducing of shunting equations that effectively handles the SMC's chattering issue and makes control activities more consistent and practical, strictly speaking, the two-phase characteristic of SMC scheme is no longer existing, i.e., reaching phase and sliding mode phase, but the resultant system is of a desirable cascaded interconnection with modular architecture consisted of several subsystems, which actually results from the design procedure of sliding mode control. Such a modularity property is very helpful and beneficial for the control system analysis and parameter tuning in practice.
\end{remark}

\begin{remark}\label{remark5}
It is observed from equations \eqref{eq25}--\eqref{eq27} that the order of the system is extended under the proposed shunting model based SMC scheme, from which we may say that our methodology can be regarded as a variant of the higher order sliding mode control (HOSMC) techniques. Generally speaking, in conventional HOSMC scheme an integral standard form is typically used to extend the relative degree of the system with the aims to smoothen the control inputs. Nonetheless, such an approach cannot totally render a chattering-free behavior, as the switching control law is still applied to satisfy the reaching condition. Furthermore, in order to synthesize such a controller, the assumption of bounded time derivative of disturbances is required, thus leading to a more conservative design. In a stark contrast, bioinspired HOSMC scheme completely gets rid of the chattering problem and doesn't rely on such an assumption. Most importantly, due to the filtering properties of the shunting model, a more robust behavior can be obtained with respect to the bounded disturbances as well as the noises without using the high-gain control.
\end{remark}

\section{SIMULATION RESULTS}\label{s4}
To validate the effectiveness of proposed formation control protocol, simulation experiments are conducted, in which four underwater vehicles steered by their own onboard controllers are employed to track a prescribed trajectory in 3-dimensional space, and meanwhile, a quadrilateral formation pattern is required to be maintained for the whole group of AUVs. { Moreover, the comparison between conventional sliding model control and proposed bioinspired control is made to show the practical improvements in terms of the smoothness of control activities, chattering suppression and disturbance rejections.}

\begin{figure}[!t]
\centering
\includegraphics[width=3.0in]{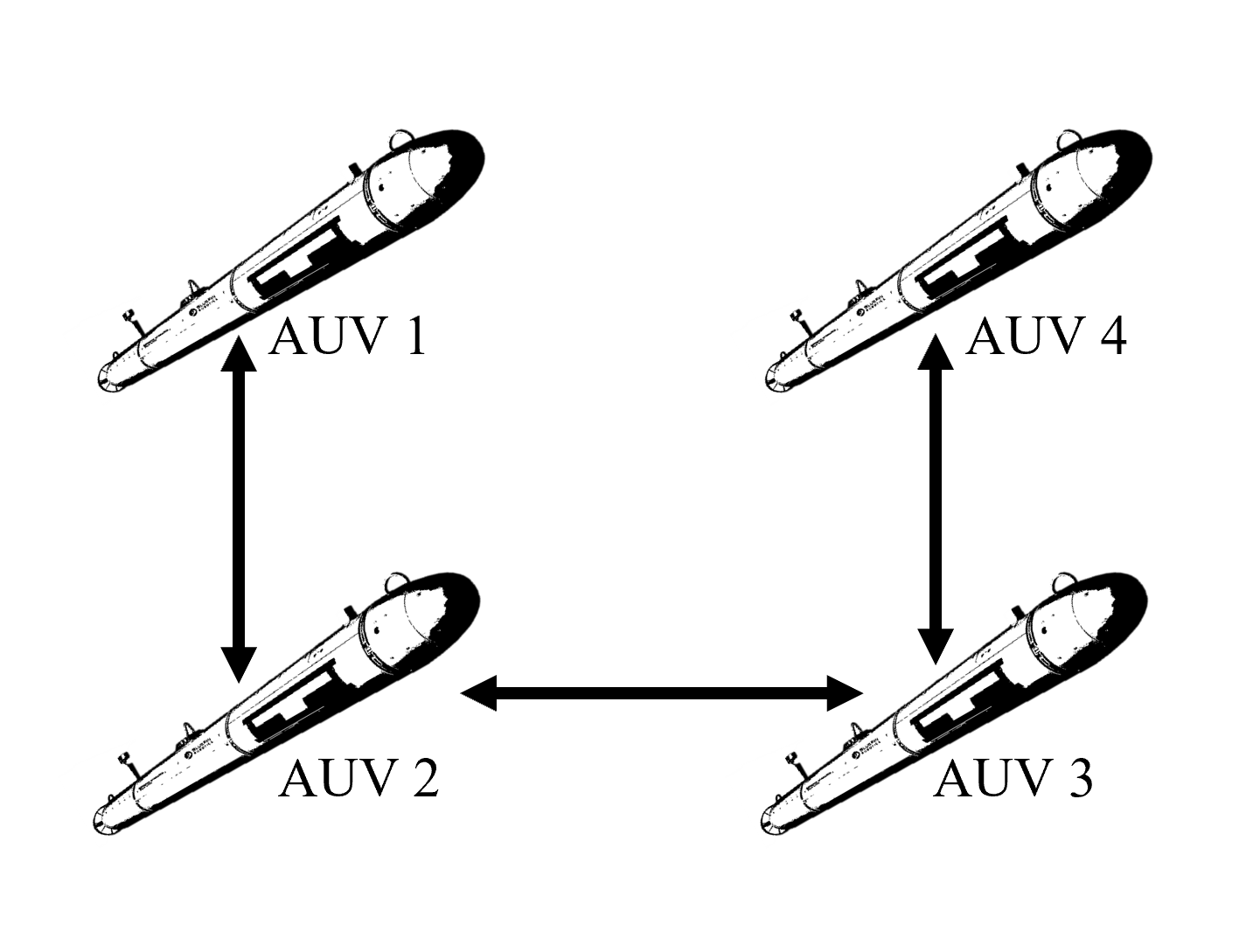}
\caption{The communication topology graph for 4 AUVs formation control.}
\label{fig_2}
\end{figure}

{Each vehicle’s dynamics used in simulations is described by \eqref{eq1}, and the corresponding system and hydrodynamic parameters are selected as follows with international units}: $m_i =20$, $I_{x,i}=20$, $I_{y,i}=30$, $I_{z,i}=35$, $\beta _{vx,i} = -8$, $\beta _{vy,i} = -10$, $\beta _{vz,i} = -9$, $\beta _{\dot vx,i} = -7$, $\beta _{\dot vy,i} = -8$, $\beta _{\dot vz,i} = -6$, $\beta _{\omega x,i} = -0.2$, $\beta _{\omega y,i} = -0.25$, $\beta _{\omega z,i} = -0.15$, $\beta _{\dot \omega x,i} = -20$, $\beta _{\dot \omega y,i} = -30$, $\beta _{\dot \omega z,i} = -35$, $\left( {i \in \left\{ {1,2,3,4} \right\}}  \right)$. The information exchange among the vehicles is illustrated in Fig. \ref{fig_2}, and the weights of the undirected edges are given as ${a_{12}} = {a_{21}} = {a_{23}} = {a_{23}} = {a_{34}} = {a_{43}} = 1$, and it is assumed that only the AUV $1$ can assess the information of desired trajectory, i.e., ${b_1} = 1$. The desired trajectory is defined as $\eta _1^d\left( t \right) = {\left[ {30 - 30{e^{ - t}},5t,2t} \right]^{\rm T}}$, and each AUV’s posture in group is required to align with $\eta _2^d\left( t \right) = {\left[ {0,0,0} \right]^{\rm T}}$. The relative distances between AUVs that are designed to form the desired quadrilateral formation pattern are given by ${\delta _{12}} = {\left[ {0,10,0} \right]^{\rm T}}$, ${\delta _{21}} = {\left[ {0, - 10,0} \right]^{\rm T}}$, ${\delta _{23}} = {\left[ { - 10,0,0} \right]^{\rm T}}$, ${\delta _{32}} = {\left[ {10,0,0} \right]^{\rm T}}$, ${\delta _{34}} = {\left[ {0, - 10,0} \right]^{\rm T}}$ and ${\delta _{43}} = {\left[ {0,10,0} \right]^{\rm T}}$. The initial positions of four vehicles are set as follows: ${\eta _1} = {\left[ {3,3,3,0.3,0,0.2} \right]^{\rm T}}$, ${\eta _2} = {\left[ {2.5,3.5,3,0.2,0,0.25} \right]^{\rm T}}$, ${\eta _3} = {\left[ {2,3,3,0.3,0,0.2} \right]^{\rm T}}$ and ${\eta _4} = {\left[ {3,3,2,0.3,0,0.2} \right]^{\rm T}}$; the initial translational and rotational velocities of AUVs are all set to ${v_i} = {\mathbf{0}_{6 \times 1}},\left( {i \in \left\{ {1,2,3,4} \right\}} \right)$. The periodic external disturbances to simulate the impacts of wind, waves as well as ocean currents are given by $d_i = \left[ {{2.5\sin \left( t \right),2.5\cos \left( t \right),2.5\sin \left( t \right)}}  \right.$, $\left. {{0.5\sin \left( t \right),0.5\cos \left( t \right),0.5\sin \left( t \right)}}  \right]$, $\left( {i \in \left\{ {1,2,3,4} \right\}} \right)$. The parameters of sliding model controller are chosen as ${k_1} = \text{diag}\left\{ {10,10,10,10} \right\}$, ${\beta _0} = \text{diag}\left\{ {100,100,100,100} \right\}$, and the bioinspired controller’s parameters are selected as ${A_s} = \text{diag}\left\{ {100,100,100,100} \right\}$, ${B_s} = \text{diag}\left\{ {30,30,30,30} \right\}$, ${D_s} = \text{diag}\left\{ {30,30,30,30} \right\}$, ${k_1} = \text{diag}\left\{ {10,10,10,10} \right\}$ and ${\beta _0} = \text{diag}\left\{ {10,10,10,10} \right\}$. It is worth noting that considering the practical saturation issue that quite often occurs in real-world actuators, the amplitudes of controller’s outputs are artificially confined into the interval $\left[ { - 300,300} \right]$. Moreover, because of the high-frequency switching nature of the conventional sliding mode control that makes the control signals oscillate extremely fast, to moderate this phenomenon saturation functions are used in simulation to take place of the sign functions that were employed in foregoing derivation of SMC controller. 
\begin{figure}[!htb]
\centering
\includegraphics[trim=250 150 150 100, clip,width=2.8in, height=2.3in]{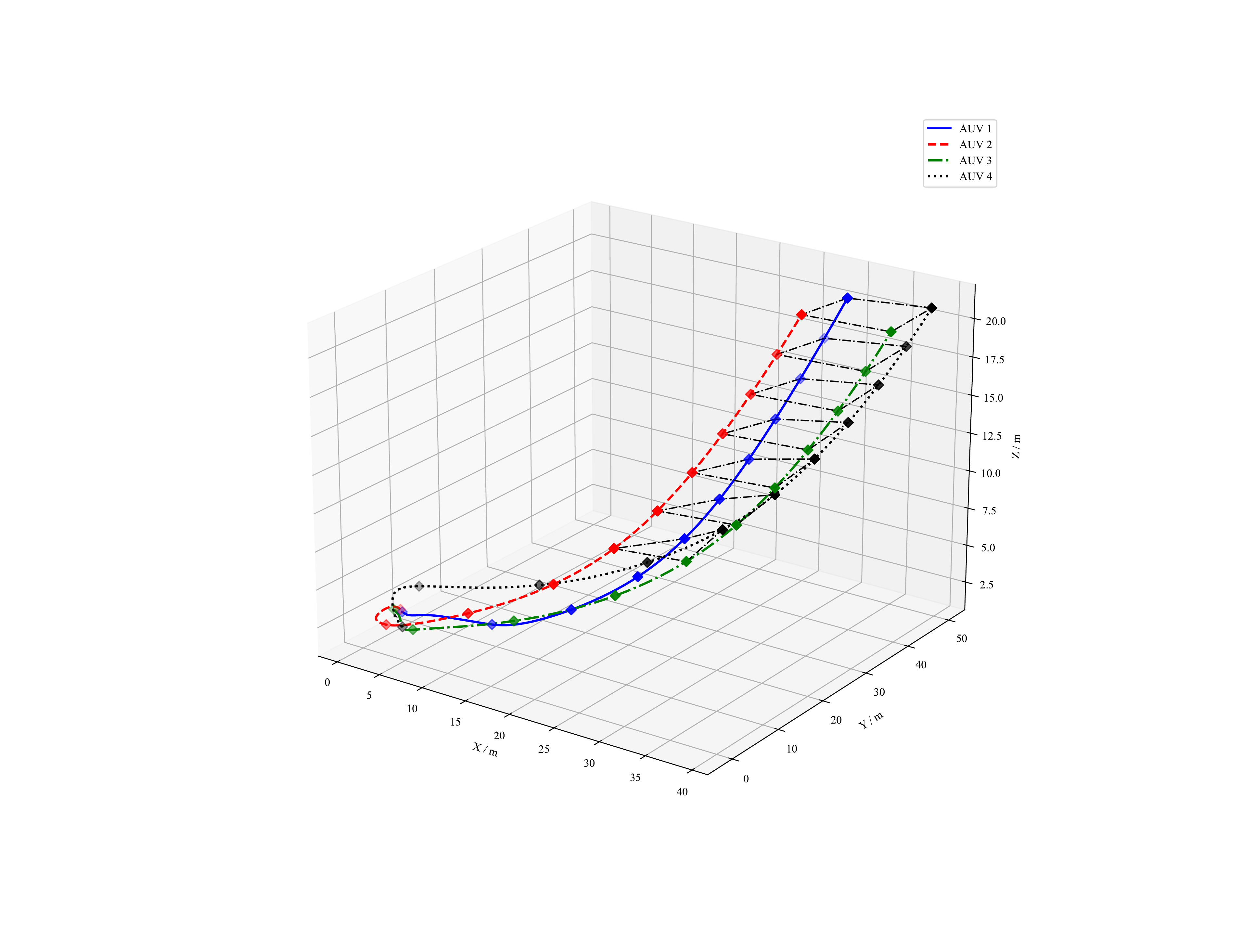}
\caption{The actual tracking trajectories and formation shape of four AUVs under bioinspired control approach.}
\label{fig_3}
\end{figure}

\begin{figure}[!htb]
\centering
\includegraphics[width=3.0in]{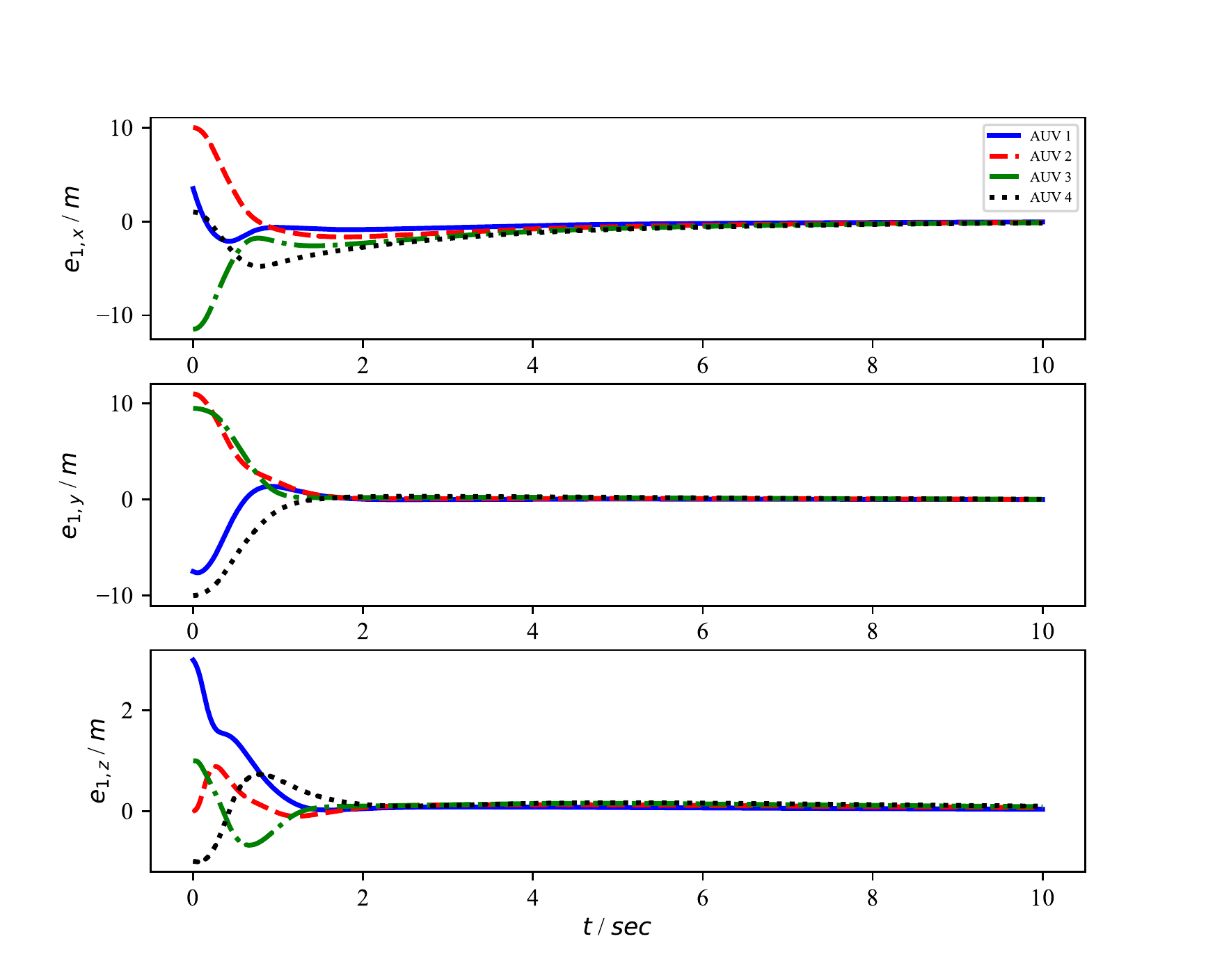}
\caption{The consensus position error of AUVs under bioinspired control approach.}
\label{fig_4}
\end{figure}

\begin{figure}[!htb]
\centering
\includegraphics[width = 3.0in]{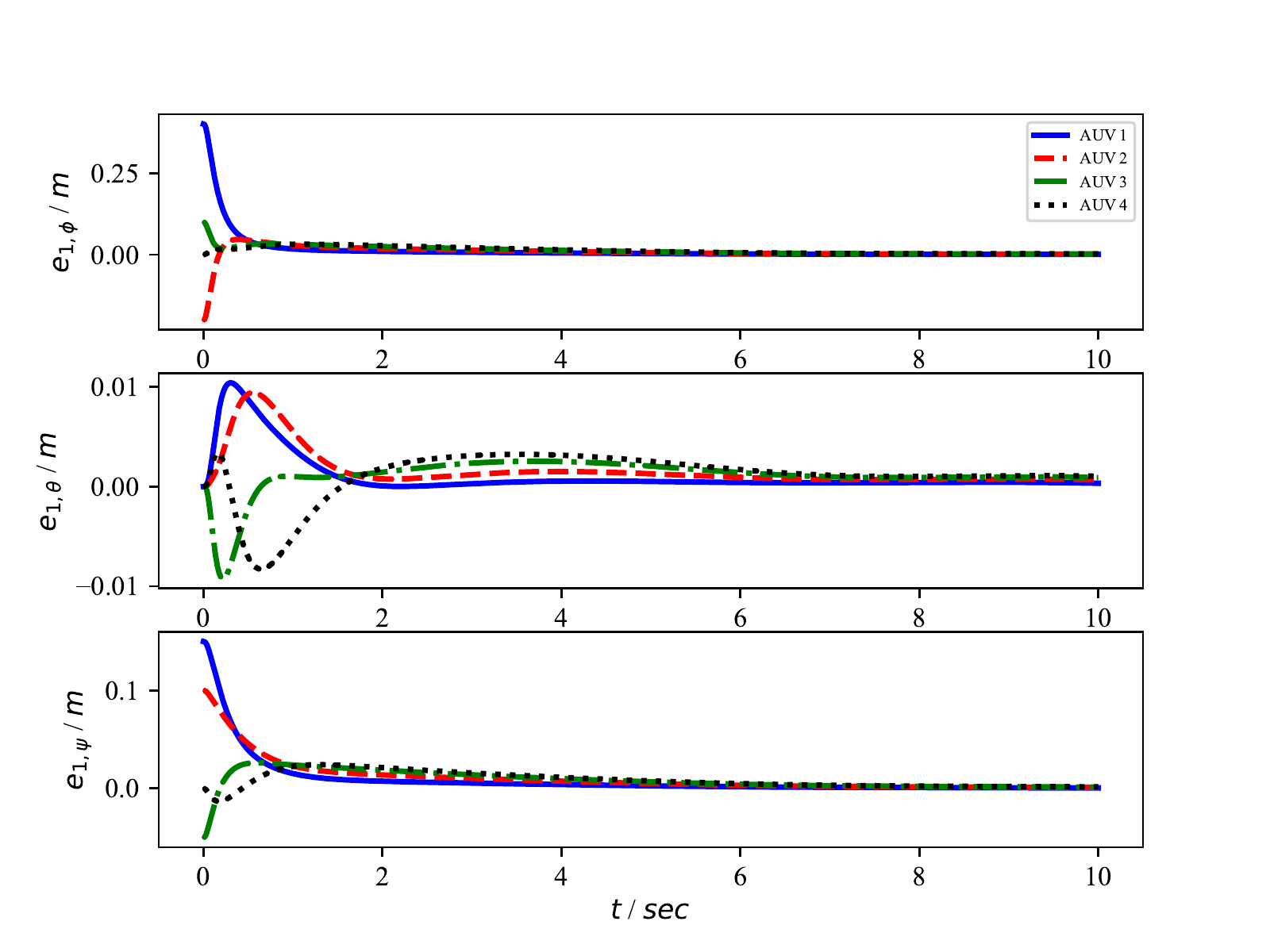}
\caption{The consensus attitude error of AUVs under bioinspired control approach.}
\label{fig_5}
\end{figure}

\begin{figure}[!htb]
\centering
\includegraphics[width=3.0in]{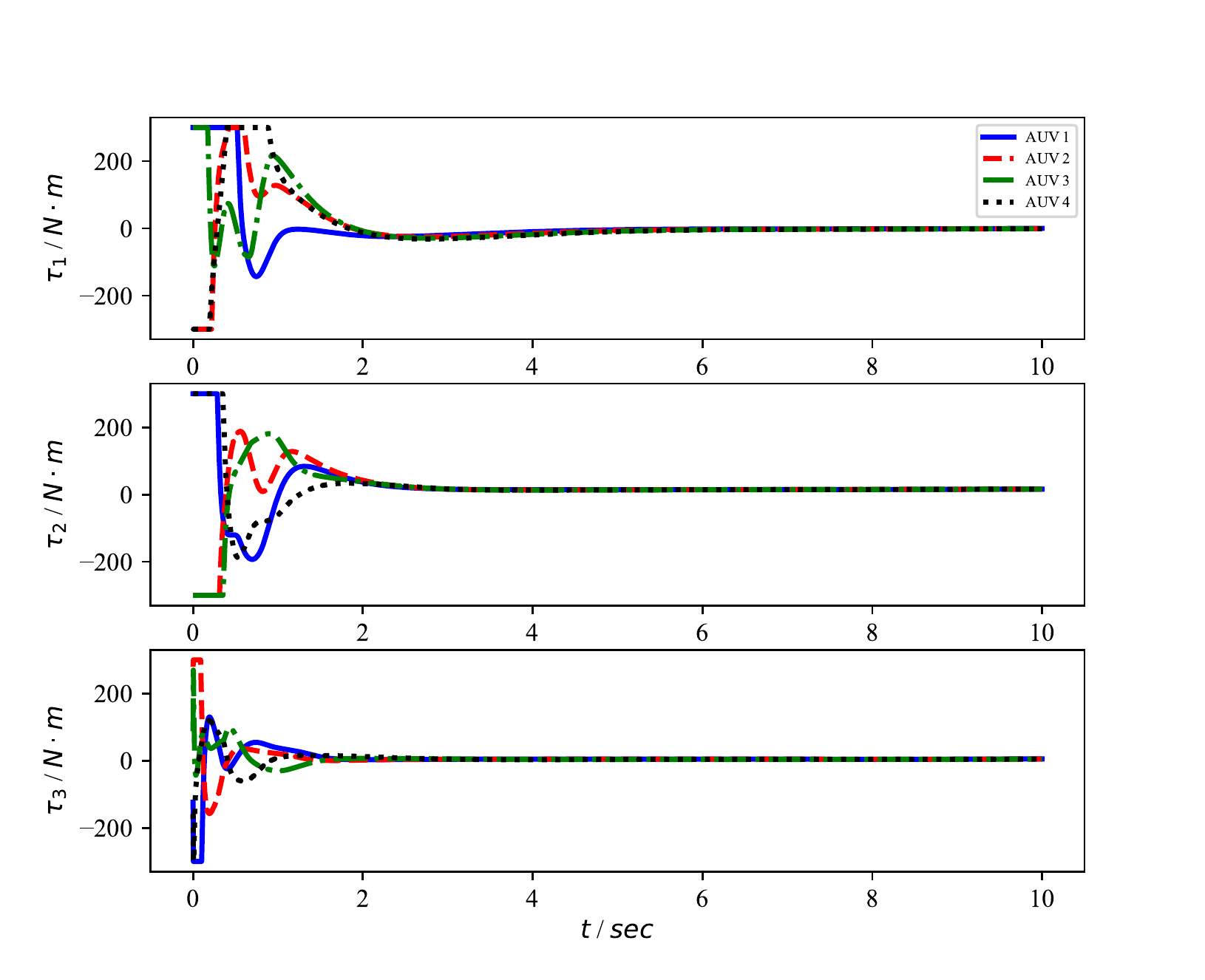}
\caption{Translational control inputs of AUVs under bioinspired control approach.}
\label{fig_6}
\end{figure}

\begin{figure}[!htb]
\centering
\includegraphics[width = 3.0in]{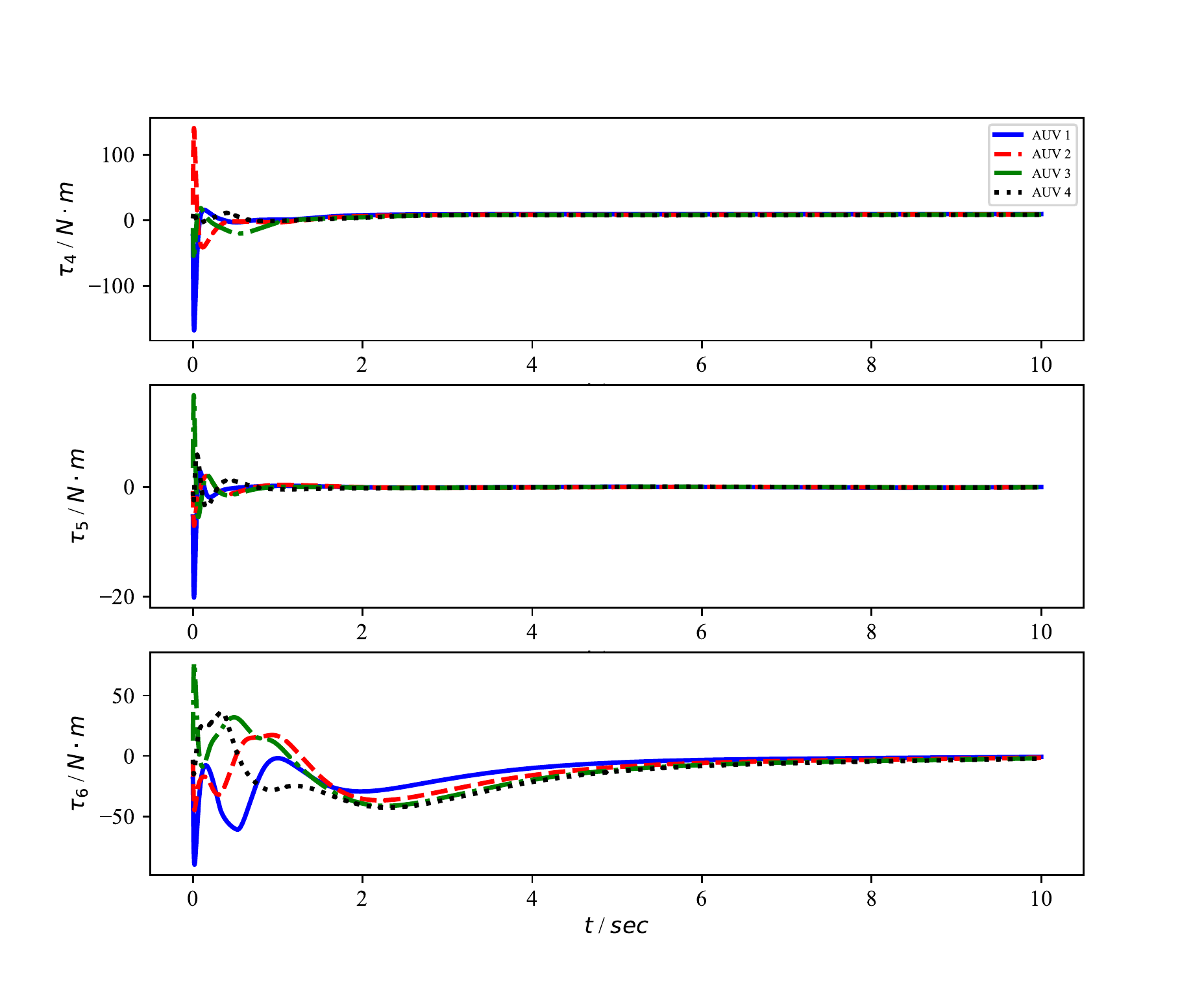}
\caption{Rotational control inputs of AUVs under bioinspired control approach.}
\label{fig_7}
\end{figure}
 The formation performance using the proposed distributed bioinspired control protocol is shown in Fig.s \ref{fig_3}-\ref{fig_7}. Fig. \ref{fig_3} shows that the control objectives are perfectly achieved, i.e., under the proposed controller overall multi-AUV system can form the desired quadrilateral formation pattern, and at same time the prescribed trajectory is followed by the group of AUVs even in the presence of external disturbances. It can also be seen from Fig.s \ref{fig_4} and \ref{fig_5} that the corresponding consensus tracking errors of all AUVs in each degree of freedom are convergent uniformly asymptotically to a very small region of the origin, which also indicates that the resulting system has good robustness property. In addition to that, it can be observed that the consensus tracking errors of each vehicle almost tend to zero simultaneously, suggesting that a good coordination performance is also achieved. Fig.s \ref{fig_6} and \ref{fig_7} show the control signals generated by the bioinspired controller, from which one can find that the chattering issue is totally disappeared in proposed control scheme due to employment of shunting model, and the control actions, thus, become much smoother, which is far more critical for the practical applications. It should be noted that because of the transient response of the shunting model some control saturations can be occurred at the beginning of time, but few moments later the control behaviors soon get into normal.
 
\begin{figure}[!t]
\centering
\includegraphics[trim=250 20 150 20, clip,width=3.0in, height=3.0in]{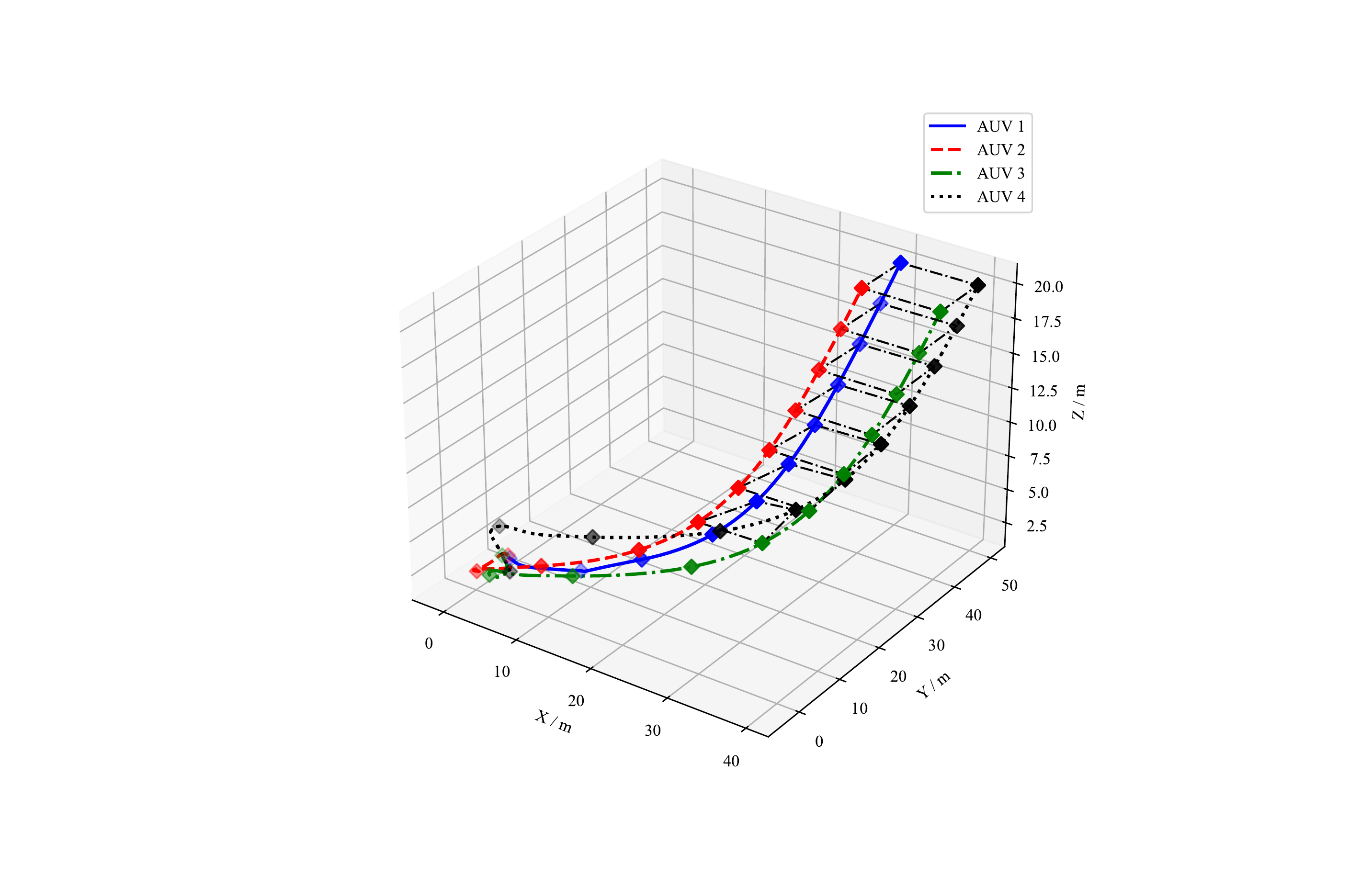}
\caption{The actual tracking trajectories and formation shape of four AUVs under sliding model control.}
\label{fig_8}
\end{figure}

\begin{figure}[!htb]
\centering
\includegraphics[width=3.0in]{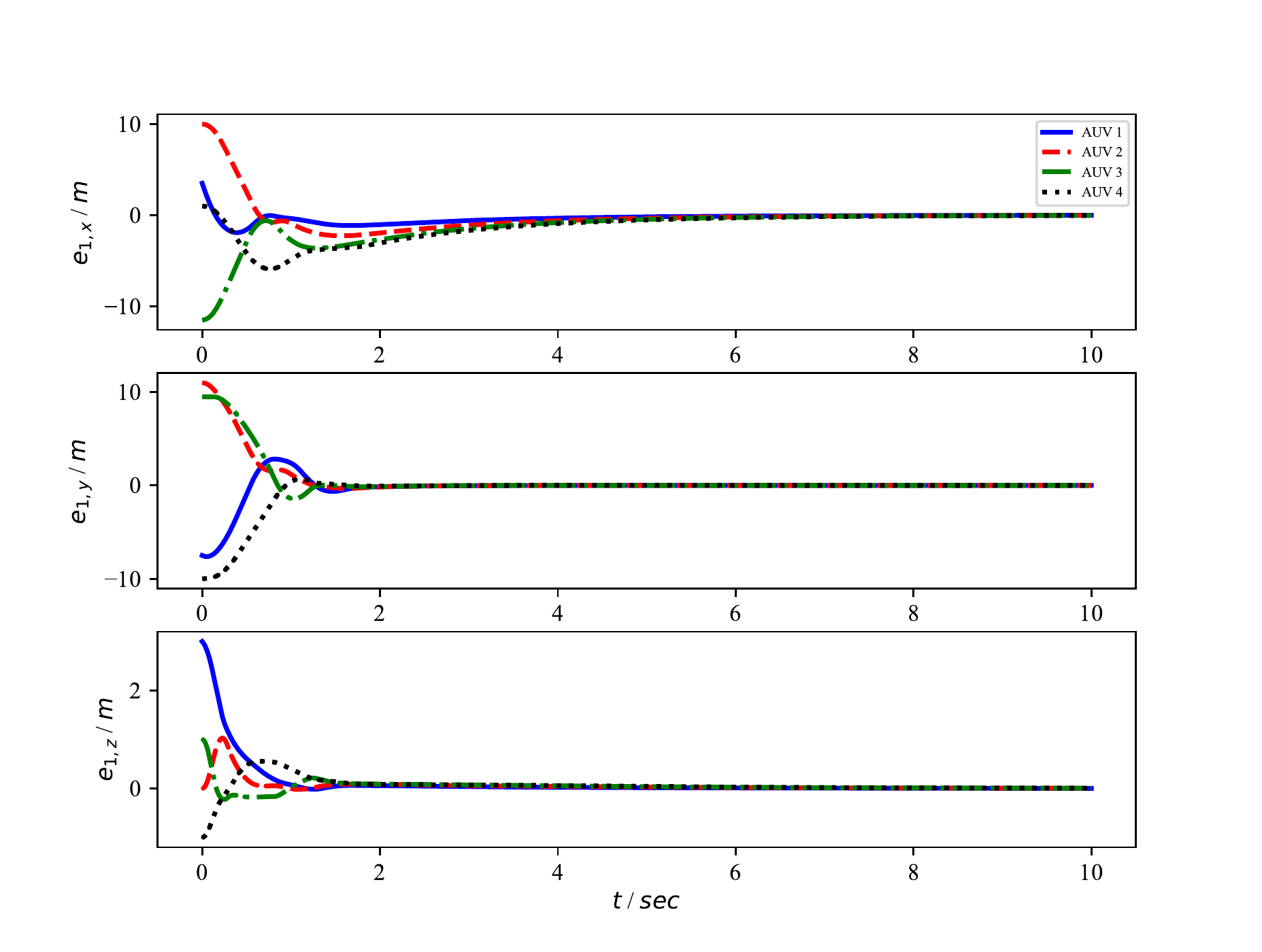}
\caption{The consensus position error of AUVs under sliding model control.}
\label{fig_9}
\end{figure}

\begin{figure}[!htb]
\centering
\includegraphics[width = 3.0in]{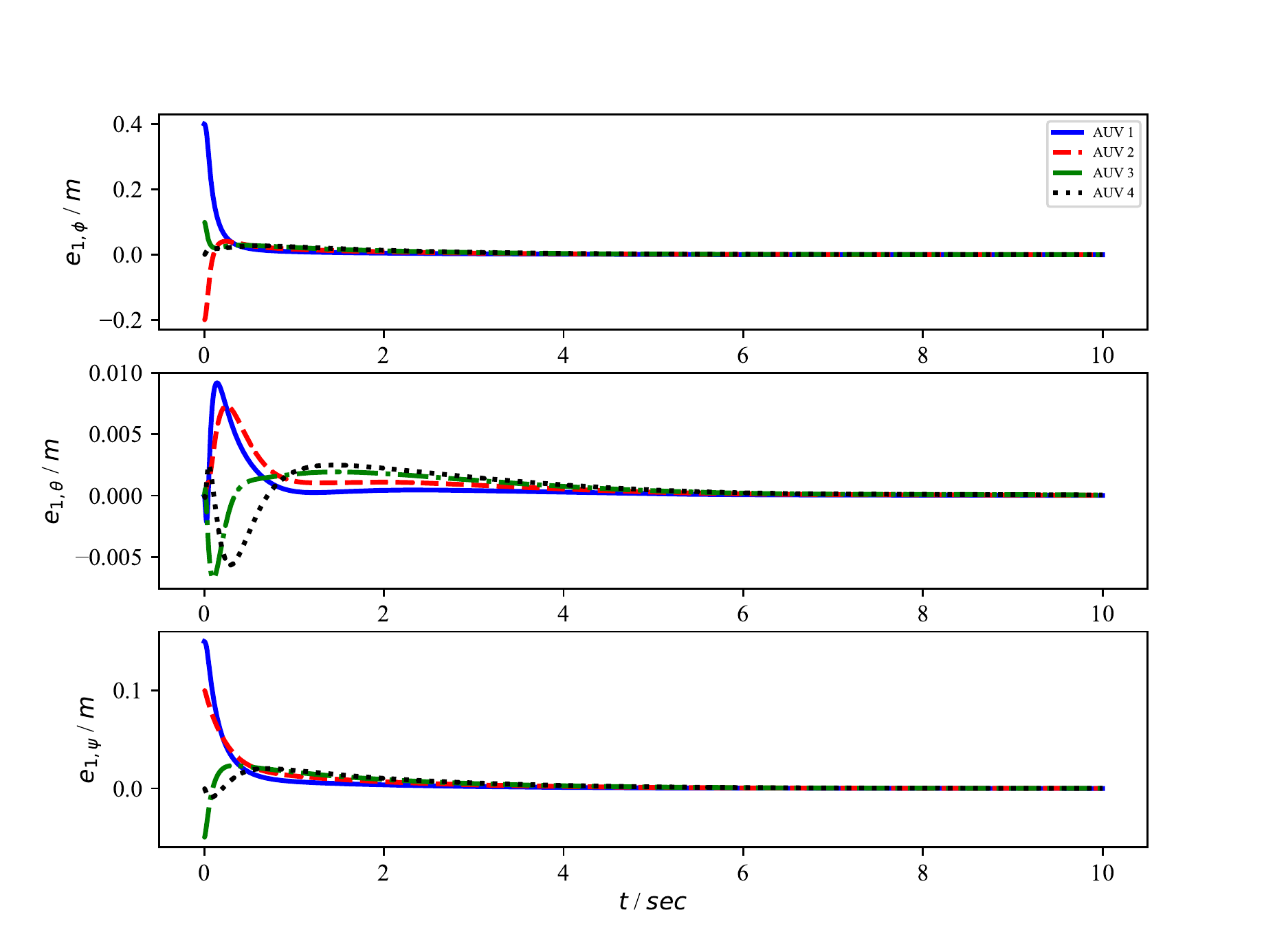}
\caption{The consensus attitude error of AUVs under sliding model control.}
\label{fig_10}
\end{figure}

\begin{figure}[!htb]
\centering
\includegraphics[width=3.0in]{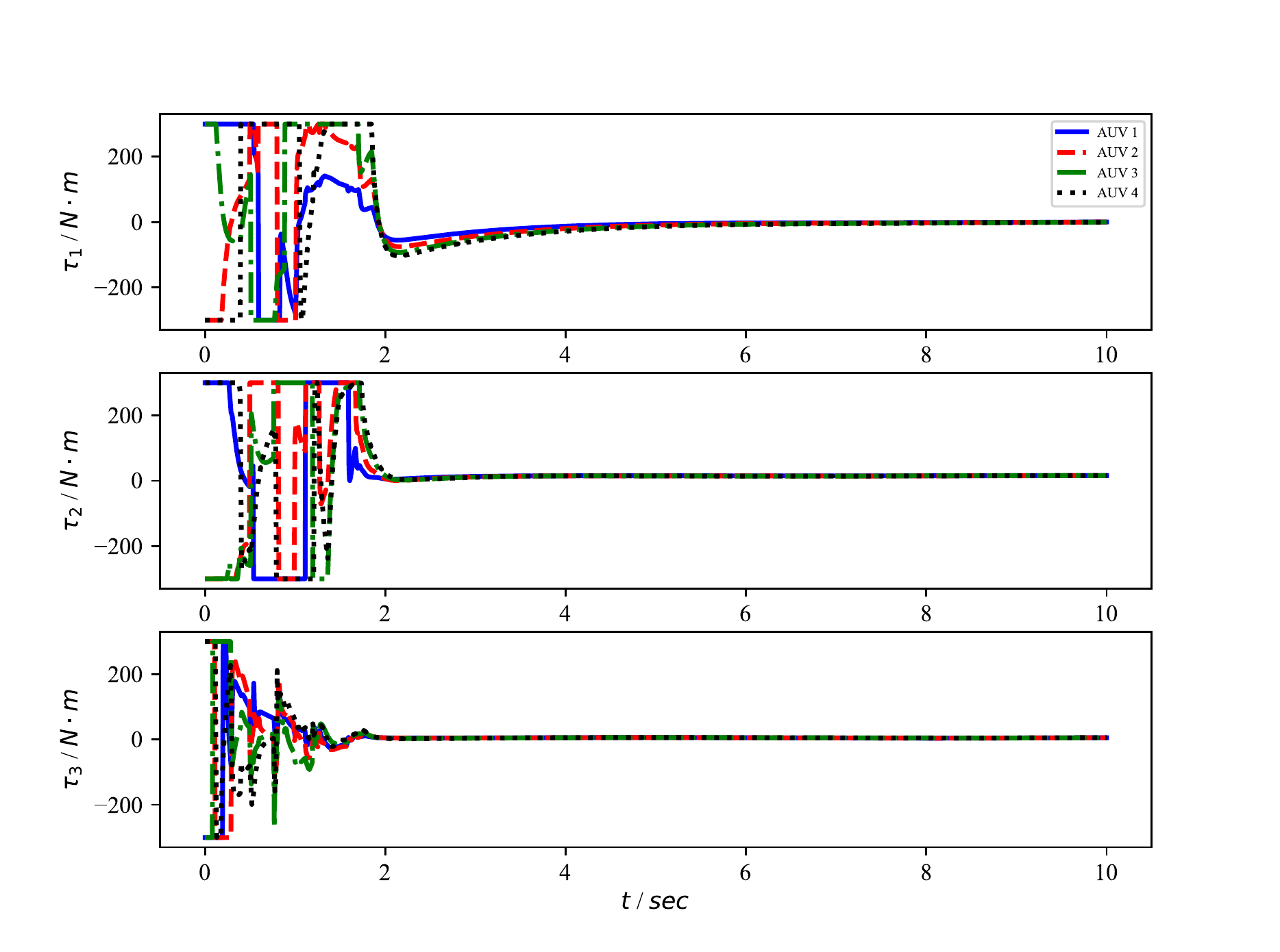}
\caption{Translational control inputs of AUVs under sliding model control.}
\label{fig_11}
\end{figure}

\begin{figure}[!htb]
\centering
\includegraphics[width = 3.0in]{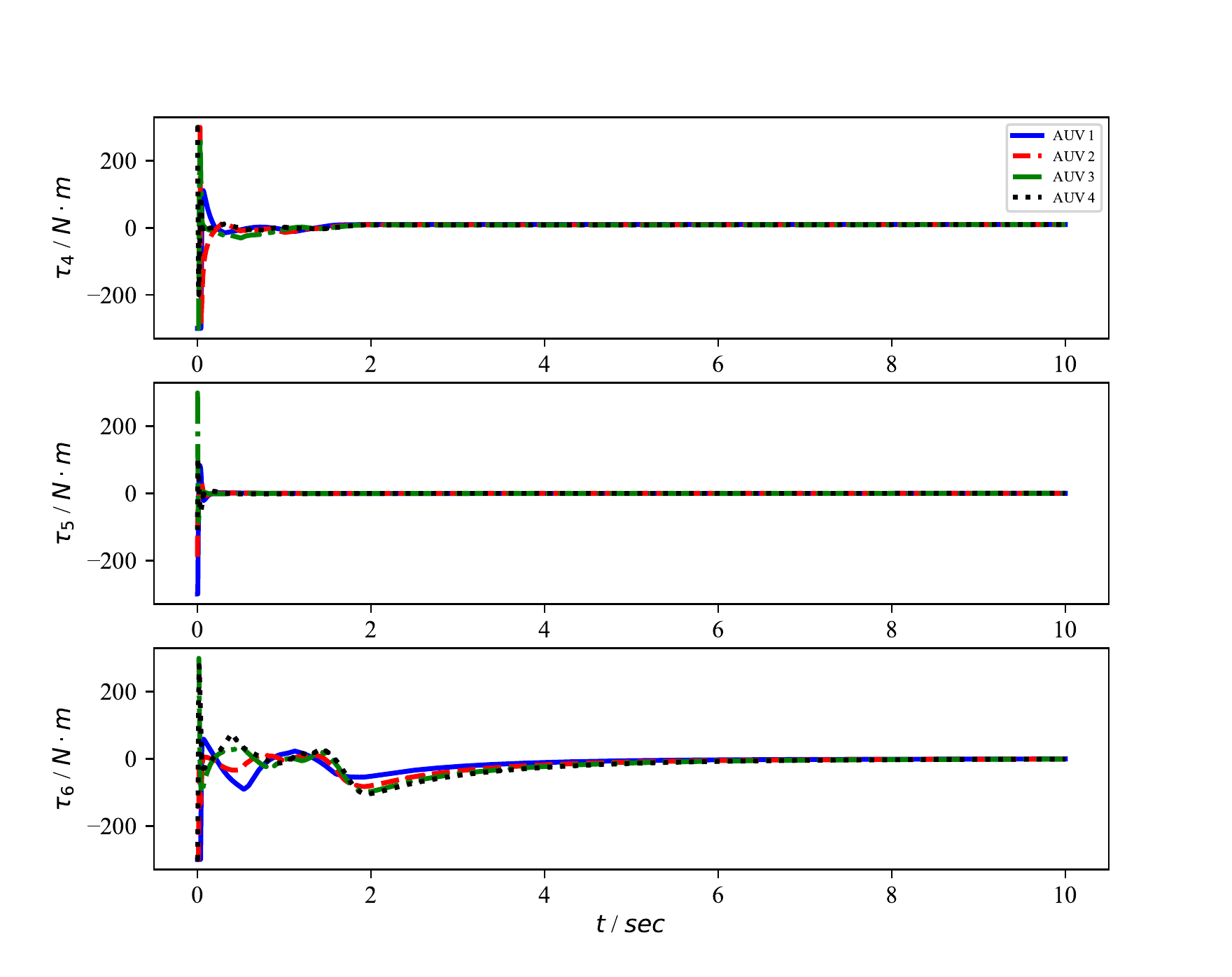}
\caption{Rotational control inputs of AUVs under sliding model control.}
\label{fig_12}
\end{figure}

The next case as the comparison shows the formation performance using the conventional sliding mode control protocol. It can be seen from Fig.s \ref{fig_8}-\ref{fig_10} that the SMC approach can also achieve the prescribed control objectives. However, the corresponding control signals as shown in Fig.s \ref{fig_11} and \ref{fig_12} exhibit apparent high-frequency oscillations and the non-smoothness, even though the saturation function is employed in place of the classic sign function which undoubtedly will further exacerbate this phenomenon. As an immediate consequence, behavior of sliding mode controller is almost unable to be realized by the actual actuators, since any physical device has its own response frequency and bandwidth. Besides, it may also excite the system’s high frequency unmodeled dynamics. Therefore, the actual formation performance by applying such control law may not be as good as the simulation results; instead, it may even render the closed-loop system unstable. In an obvious contrast, it can be seen from Fig.s \ref{fig_6} and \ref{fig_7} that the proposed bioinspired SMC scheme completely overcomes these drawbacks; that is, there is no high-frequency switching behavior arising in the control process, and the control signals are much smoother than the conventional SMC approach.

\section{CONCLUSION}\label{s5}
{The formation tracking control of multiple AUV systems in 3-dimensional space subject to the hydrodynamic parameter uncertainties and unknown environmental disturbances is addressed in this paper, where the overall system is not only required to maintain a specified formation pattern, but is aimed to follow a desired trajectory as a whole. In order to obtain a good coordination performance in shape forming, consensus problem is also considered between neighbor vehicles. To this end, a novel distributed bioinspired sliding mode control protocol is proposed based on the graph theory. In particular, due to the favorable dynamic characteristic of the shunting model, the chattering phenomena are completely avoided in the present SMC scheme, the amplitude of the control is able to be bounded, and the generated control actions are more consistent and smoother compared to the conventional one. Most importantly, owing to the filtering nature of the shunting equation, more robust behavior can be obtained with respect to the bounded lumped disturbance as well as the noises without needing the large control efforts. Moreover, the stability analysis of the resulting closed-loop system is proved using the Lyapunov direct method. Finally, several simulation experiments are carried out to demonstrate the efficiency and effectiveness of proposed distributed formation control protocol. In future work, more practical constraints on multiple AUV formation systems need to be further taken into account and addressed, such as the impacts of the communication imperfections (e.g., time delays, package loss, and even misleading data) as well as the nonholonomic kinematics.}

% \section*{Acknowledgments}
% This should be a simple paragraph before the References to thank those individuals and institutions who have supported your work on this article.

% {\appendix[Proof of the Zonklar Equations]
% Use $\backslash${\tt{appendix}} if you have a single appendix:
% Do not use $\backslash${\tt{section}} anymore after $\backslash${\tt{appendix}}, only $\backslash${\tt{section*}}.
% If you have multiple appendixes use $\backslash${\tt{appendices}} then use $\backslash${\tt{section}} to start each appendix.
% You must declare a $\backslash${\tt{section}} before using any $\backslash${\tt{subsection}} or using $\backslash${\tt{label}} ($\backslash${\tt{appendices}} by itself
%  starts a section numbered zero.)}

% %{\appendices
% %\section*{Proof of the First Zonklar Equation}
% %Appendix one text goes here.
% % You can choose not to have a title for an appendix if you want by leaving the argument blank
% %\section*{Proof of the Second Zonklar Equation}
% %Appendix two text goes here.}

% \section{References Section}
% You can use a bibliography generated by BibTeX as a .bbl file.
%  BibTeX documentation can be easily obtained at:
%  http://mirror.ctan.org/biblio/bibtex/contrib/doc/
%  The IEEEtran BibTeX style support page is:
%  http://www.michaelshell.org/tex/ieeetran/bibtex/
 
 % argument is your BibTeX string definitions and bibliography database(s)
\bibliographystyle{IEEEtran}
%\bibliographystyle{plain}
% argument is your BibTeX string definitions and bibliography database(s)
\bibliography{reference}
%\input{output.bbl}
%\bibliography{output.bbl}

\newpage
\vspace{11pt}

\begin{IEEEbiography}[{\includegraphics[width=1in,height=1.25in,clip,keepaspectratio]{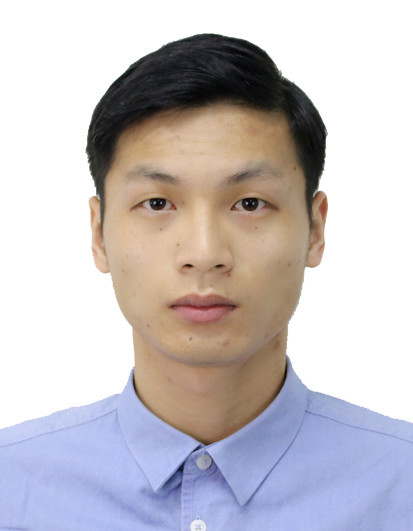}}]{Tao Yan}
(S’22) received the B.Sc. degree from the
North China Institute of Aerospace Engineering, Langfang, China, in 2016; the
M.Sc. degree from the Zhejiang University of Technology,
Hangzhou, China, in 2020. He is currently pursuing
his Ph.D. degree at the University of Guelph,
ON, Canada. His research interests include the 
intelligent control, distributed control and optimization, and networked underwater vessel systems.
\end{IEEEbiography}

\vspace{11pt}

\begin{IEEEbiography}[{\includegraphics[width=1in,height=1.25in,clip,keepaspectratio]{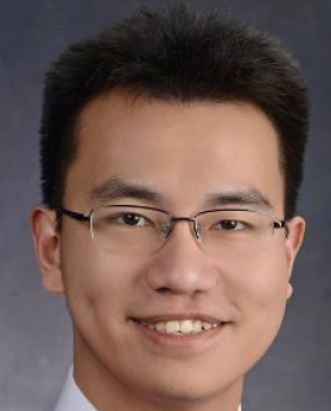}}]{Zhe Xu}
(S’20) received the
B.Eng. degree in mechanical engineering and the
M.A.Sc. degree in engineering systems and computing
from the University of Guelph, in 2018 and
2019, respectively, where he is currently pursuing
the Ph.D. degree in engineering systems and
computing with the School of Engineering, under
the supervision of Prof. S. X. Yang. His research
interests include tracking control, estimation theory,
robotics, and intelligent systems.
\end{IEEEbiography}

\vspace{11pt}

\begin{IEEEbiography}[{\includegraphics[width=1in,height=1.25in,clip,keepaspectratio]{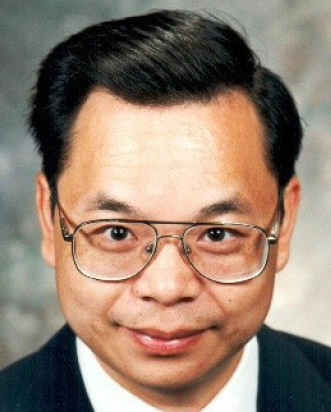}}]{Simon X. Yang}
(S’97–M’99–SM’08) received the
B.Sc. degree in engineering physics from Beijing
University, Beijing, China, in 1987, the first of
two M.Sc. degrees in biophysics from the Chinese
Academy of Sciences, Beijing, in 1990, the second
M.Sc. degree in electrical engineering from the
University of Houston, Houston, TX, in 1996, and
the Ph.D. degree in electrical and computer engineering
from the University of Alberta, Edmonton,
AB, Canada, in 1999. Dr. Yang is currently a
Professor and the Head of the Advanced Robotics
and Intelligent Systems Laboratory at the University of Guelph, Guelph, ON,
Canada. His research interests include robotics, intelligent systems, sensors
and multi-sensor fusion, wireless sensor networks, control systems, machine
learning, fuzzy systems, and computational neuroscience. Prof. Yang he
has been very active in professional activities. He serves as the Editor-in-
Chief of International Journal of Robotics and Automation, and an Associate
Editor of IEEE Transactions on Cybernetics, IEEE Transactions of Artificial
Intelligence, and several other journals. He has involved in the organization
of many international conferences.
\end{IEEEbiography}

\vspace{11pt}

\vfill

\end{document}